\documentclass[useAMS,usenatbib,usegraphicx]{mn2e}

\graphicspath{{./figures/}}

\newcommand{\D}{\Delta}              
\newcommand{\x}{\mathbf{\times}}     
\newcommand{\pd}{\partial}           

\newcommand{\Uvec}[1]%
{\ensuremath{\mathbf{e}_{#1}}}  

\renewcommand{\vec}[1]%
{\ensuremath{\bmath{#1}}}       

\newcommand{\B}{\vec{B}}        
\newcommand{\E}{\vec{E}}        

\newcommand{\GJ}[1]{\ensuremath{#1_\textrm{\tiny GJ}}}    
\newcommand{\Rgj}{\GJ{\rho}} 

\newcommand{\RPC}{\ensuremath{r_\mathrm{pc}}}     
\newcommand{\QPC}{\ensuremath{\theta_\mathrm{pc}}}

\newcommand{\Pol}[1]{\ensuremath{#1_\mathrm{pol}}}    
\newcommand{\PC}[1]{\ensuremath{#1_\mathrm{pc}}}

\newcommand{\Ud}{\ensuremath{\vec{U}_\mathrm{D}}}

\newcommand{\OmF}{\ensuremath{\Omega_\mathrm{F}}}
\newcommand{\SSp}{\ensuremath{SS^\prime}}       
\newcommand{\PsiL}{\ensuremath{\psi_\mathrm{last}}} 

\newcommand{\Wmd}{\protect{\ensuremath{|W_\mathrm{md}|}}}

\newcommand{\XL}{\ensuremath{x_0}}
\newcommand{\RLC}{\ensuremath{R_\mathrm{LC}}}     
\newcommand{\RNS}{\ensuremath{R_\mathrm{NS}}}      

\newcommand{\XNS}{\ensuremath{x_\textrm{\tiny NS}}}
\newcommand{\ZNS}{\ensuremath{z_\textrm{\tiny NS}}}
\newcommand{\XMax}{\ensuremath{x_\mathrm{max}}}
\newcommand{\ZMax}{\ensuremath{z_\mathrm{max}}}

\title%
{On the force-free magnetosphere of aligned rotator}
\author[A.~N.~Timokhin]{%
A.~N.~Timokhin\thanks{E-mail: atim@sai.msu.ru}\\
Sternberg Astronomical Institute, Universitetskij pr. 13, 119992 Mocsow, Russia}

\begin{document}

\date{Accepted Received ; in original form }

\pagerange{\pageref{firstpage}--\pageref{lastpage}} \pubyear{}

\maketitle

\label{firstpage}

\begin{abstract}
  We investigate in details properties of stationary force-free
  magnetosphere of aligned rotator assuming the last closed field line
  lying in equatorial plane at large distances from pulsar.  The
  pulsar equations is solved numerically using multigrid code with
  high numerical resolution, physical properties of the magnetosphere
  are obtained with high accuracy. We found a set of solutions with
  different sizes of the closed magnetic field line zone and verify
  the applicability of the force-free approximation. We discuss the
  role of electron-positron cascades in supporting of the force-free
  magnetosphere and argue that the closed field line zone should grow
  with time slower than the light cylinder. This yield the pulsar
  breaking index less than 3. It is shown, that models of aligned
  rotator magnetosphere with widely accepted configuration of magnetic
  field, like one considered in this paper, have serious difficulties.
  We discuss solutions of this problem and argue that in any case
  pulsar energy losses should evolve with time differently than
  predicted by the magnetodipolar formula.
\end{abstract}


\section{Introduction}
\label{sec:introduction}

Since the first works on pulsar magnetosphere, a stationary force-free
magnetosphere of aligned rotator is considered as an underlying model
for the real pulsar magnetosphere for more than 30 years. Despite its
degenerated character (such ``pulsar'' even does not pulse), it is
believed to reproduce qualitatively all main properties of the real
pulsar magnetosphere. For near-aligned pulsars it should give even an
adequate detailed description. The structure of aligned rotator's
force-free magnetosphere can be described by solution of a single
scalar non-linear PDE, the so-called ``pulsar equation'', derived by
\citet{Michel73,ScharlemannWagoner73,Okamoto74}.  This is an equation
for the flux of poloidal magnetic field. All other physical quantities
describing the magnetosphere are related to the flux function $\Psi$,
poloidal current $J$ and angular velocity of magnetosphere's rotation
$\Omega$ by algebraic relations.

Analytical solution of this equation with non-zero poloidal current
seems exist only for the split-monopole configuration of the magnetic
field \citep{MichelBook} and for a slightly perturbed split monopole
\citep{Beskin/Kuznetsova/98}. For dipole magnetic field analytical
solution for the case of zero poloidal current has been found
\citep{Michel73:b,Mestel/Wang79}, but this solution is valid only
inside the light cylinder (LC).  There were several works dedicated to
solution of linearised pulsar equation, where poloidal current and
angular velocity were assumed to be proportional to magnetic flux
function, what made the equation linear, but they did not lead to
construction of a consistent model of aligned rotator magnetosphere
\citep[see e.g.][]{Beskin/83,Lyubar90,Beskin/Malyshkin98}.

The first attempt to solve this equation numerically was made by
\citet{CKF}, hereafter CKF. They have shown for the first time, that
there exists a \emph{self-consistent} solution with dipole magnetic
field geometry near the NS and magnetic field lines smoothly passing
through the light cylinder. In that work the position of the null
point%
\footnote{the point where the last closed field line intersects the
  equatorial plane} %
was fixed at the light cylinder and the question about applicability
of the force-free approximation have been not investigated.  Energy
losses of the aligned rotator for CKF solution have been calculated by
\citet{Gruzinov:PSR}. \citet{Goodwin/04} have studied this problem
more deeply, namely they have searched for solution of the pulsar
equation when the position of the null point is not fixed at the LC,
but lies at different positions inside the light cylinder. For any
position of the null point they obtained solutions, smoothly passing
the light cylinder, but like CKF they have not studied physical
properties of obtained solutions (e.g.  energy losses, applicability
of the force-free approximations etc.). Their model, however, seems to
be artificial, because they assumed non-zero pressure in the closed
field line zone, what implies continuous energy injection into the
closed field line domain.  Recently \citet{Contopoulos05} addressed
the case when the plasma rotation frequency in the open field line
domain is different from the rotation frequency of the NS. It was
shown, that there exist an unique solution of the pulsar equation for
arbitrary plasma rotation frequency, although a rather simple case
when the plasma rotation frequency is constant have been considered.
Applicability of the force-free approximations in the magnetosphere of
aligned rotator was considered in \citet{Timokhin05_astro/ph} and
\citet{Contopoulos05}, though in the latter work only for the null point
located at the LC.

Recently a different approach to the pulsar magnetosphere modelling is
being developed by \citet{Spitkovsky05_poland}, \citet{Komissarov05},
and \citet{McKinney:NS:06}. They perform \emph{time-dependent}
simulations of the pulsar magnetosphere. In \citet{Komissarov05}
aligned rotator magnetosphere was modelled using full MHD code, in
\citet{McKinney:NS:06} the same modelling was done with force-free
code.  The code of Anatoly Spitkovsky allows performing of 3-D
time-dependent simulation of the magnetosphere of inclined rotator by
solving equations of force-free MHD.  In these simulations the
\emph{existence} of the stationary force-free magnetospheric
configuration was rigorously proved for the first time.  Although this
approach presents a big step towards the modelling of the real pulsar
magnetosphere, in this paper we argue that it has serous limitation,
namely the properties of cascades supplying particles in magnetosphere
are not incorporated in these simulations.  As it will be discussed
later, cascades can set non-trivial boundary conditions on the current
density in the magnetosphere. Its incorporation in time-dependent
codes would require some efforts.

\begin{figure}
  \includegraphics[clip,width=\columnwidth]{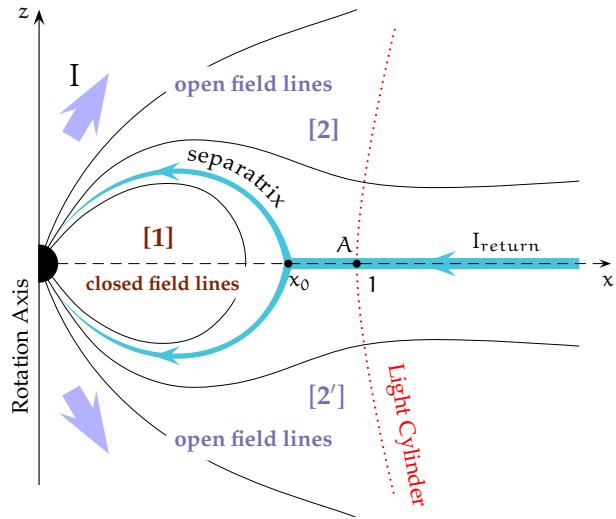} 
  \caption{Configuration of magnetic field in the magnetosphere of
    aligned rotator with Y null point -- Y-configuration. After the
    null point \XL{} the separatrix goes along the equatorial plane.
    The volume current $I$ flowing in the open filed line zones $[2]$
    and $[2^\prime]$ closes somewhere beyond the light cylinder.
    There could be a volume return current along some open field
    lines, but the largest part of it flows along the separatrix.}
  \label{fig:Topology-Y}
\end{figure}

In this work we investigate stationary problem solving the pulsar
equation numerically with high numerical resolution. We assume zero
pressure in the closed field line region (cold plasma). As in all
above mentioned works on numerical modelling of stationary aligned
rotator magnetosphere we assume a topology with the current sheet in
open field line domain flowing in the equatorial plane, i.e.
configuration with Y null point -- see Fig.~\ref{fig:Topology-Y}.
This type of magnetosphere topology had became de facto the ``standard
model'', so we study it in details and analyse its properties
regarding many aspects of electrodynamics.  Smooth solutions are
obtained for \emph{any} position of the null point inside the light
cylinder.  High numerical resolution allows accurately incorporation
of the return current flowing along the separatrix into numerical
procedure.  With the high resolution of used numerical method it was
possible to calculate accurately physical properties of the solutions
such as Goldreich-Julian charge density, magnetic field, energy losses
and pointing flux distribution etc., check applicability of the
force-free approximation and consider compatibility of the model with
models of electron-positron cascades.

Adjustment of the current density in the polar cap cascade zone of
pulsar to the global magnetospheric structure was debated already in
the first ten years after pulsar discovery \citep[see
e.g.][]{Arons79}. A concrete mechanism for the current density
adjustment was proposed by Yu.  Lyubarskij many years later, in 1992.
At that time there was no self-consistent model of pulsar
magnetosphere and detailed discussion on this subject was difficult.
Here we discuss the coupling between the polar cap cascade zone and
the rest of the magnetosphere in the frame of the self-consistent
model obtained in our simulations.  We extend the picture proposed by
\citet{Lyubar92} addressing the evolution of the current adjustment
mechanism with ageing of the pulsar. We also prove the necessity of
such mechanism and discuss it in more details in the frame of the
cascade model proposed by \citet{Arons/Scharlemann/78}.  We underline
serious difficulties of model with Y null point regarding its
compatibility with the Space Charge Limited Flow models of polar cap
cascades and briefly discuss other possible magnetospheric
configuration.

The plan of the paper is the following. In
Section~\ref{sec:pulsar-equation} important properties of the pulsar
equation are discussed. Model used in the current work and numerical
method are described in Section~\ref{sec:numerical-model}. Results of
numerical simulations are presented in
Section~\ref{sec:results-calculations}. In
section~\ref{sec:discussion} we discuss the role of polar cap cascades
for the global structure of the magnetosphere, consider in details
properties of the force-free magnetosphere with Y null point, and
highlight problems of the ``standard'' model of aligned rotator
magnetosphere. A different topology of the magnetosphere, with X null
point, is briefly discussed at the end of the section. We summarise
the most important results in Section~\ref{sec:conclusions}.

%
%

\section{The pulsar equation}
\label{sec:pulsar-equation}

\subsection{General equation}
\label{sec:general-equation}

Here we adopt the wide used assumption, that the \emph{entire}
magnetosphere of the neutron star (NS) is filled with plasma. In
some works starved magnetosphere configuration have been debated
\citep[see e.g.][]{Smith/Michel01,Petri/Heyvaerts/02}, where there are
several separated clouds of charged particles near the NS and no
particle outflow, however there are indications that such configuration is
unstable against diocotron instability
\citep{Spitkovsky/Arons02,Spitkovsky04}. Plasma in the magnetosphere
has to be non-neutral in order to screen the longitudinal (directed
along magnetic field lines) component of the electric field, induced
by rotation of the NS. In presence of the longitudinal electric field
charged particles would be accelerated and their radiation will lead
to copious electron-positron pair production in super-strong magnetic
field of pulsar \citep{Sturrock71}, what finally results in screening
of the accelerating field.

Charge density necessary for cancelling of the longitudinal electric
field, the so-called Goldreich-Julian (GJ) charge density \citep{GJ}, near the
neutron star is given by
\begin{equation}
  \Rgj \simeq - \frac{\vec{\Omega}\cdot\B}{2\pi c},
  \label{eq:rhoGJ}  
\end{equation}
where $\vec{\Omega}$ is angular velocity on neutron star rotation, $\B$
is magnetic field and $c$ is the speed of light.  Assuming that NS has
dipolar magnetic field, the ratio of the particle kinetic energy
density in the magnetosphere to the energy density of the magnetic
field at the distance $r$ can be estimated as
\begin{equation}
  \begin{array}{l}  
    \displaystyle
    \frac{\varepsilon_\mathrm{kin}}{\varepsilon_B} \sim 
    \frac{ (\Rgj/e) \; m_e c^2 \gamma}%
    { ( B^2/8\pi) }\simeq 
    \hfill\\
    \displaystyle
    \simeq  1.4\times 10^{-11}  
    P^{-1} 
    \left( \frac{\gamma}{10^7} \right)  
    \left( \frac{B_0}{10^{12}\,\mathrm{G}} \right)^{-1}
    \left( \frac{r}{\RNS} \right)^3,
  \end{array}
\end{equation}
where $e$ and $m_e$ are electron charge and mass correspondingly,
$\RNS$ -- neutron star radius, $\gamma$ -- Lorentz factor of
accelerated particles, $B_0$ magnetic field strength in Gauss near
magnetic poles of the star and $P$ -- period of pulsar rotation in
seconds. All these quantities are normalised to their typical values
in pulsars.  This ratio is small, less than 1 per cent, in
the region with the size $\sim 10^3 \, P^{-1}$ radii of the neutron
star. It could remain small even further, but here magnetic filed
deviates substantially from the NS's dipole field due to currents
flowing in the magnetosphere, and
$\varepsilon_\mathrm{kin}/\varepsilon_B$
can be estimated only after a self-consistent solution for the
magnetosphere structure is found. So, in a large domain surrounding
the neutron star, we can use force-free approximation, when particle
inertia is neglected, and equation of motion takes the form
\begin{equation}
  \rho \E + \frac1{c} [\vec{j} \x \B ] = 0.
  \label{eq:Motion:general}
\end{equation}
Hence, electric field \E{} is perpendicular to the magnetic field \B.
Charge density $\rho$ and current density \vec{j} in
eq.~(\ref{eq:Motion:general}) can be found from the Maxwell equations
(we consider stationary problem)
\begin{eqnarray}
  \label{eq:Maxwell_1}
  \nabla \cdot \E & = & 4\pi \rho
  \,,\\
  \label{eq:Maxwell_2}
  \nabla \x \B & = & \frac{4\pi}{c} \vec{j}
  \,,
\end{eqnarray} 
With help of these equations equation~(\ref{eq:Motion:general}) can be
written as
\begin{equation}
  \label{eq:Motion:EB}
  (\nabla\cdot \E)\; \E + [\nabla\x\B]\x \B = 0
  \,.
\end{equation}
In the force-free electrodynamics (FFE)%
\footnote{hereafter we use this shorter name for force-free degenerate
  electrodynamics \citep[see e.g.][and references
  there]{Komissarov02,Blandford02}}
the only possible motion of charged particles across magnetic field
lines is the drift in crossed electrical and magnetic fields with the
velocity
\begin{equation}
  \label{eq:Udrift:general}
  \Ud = c\;\frac{\E \x \B }{B^2}
  \,.
\end{equation}
Obviously $|\Ud|$ must be less than $c$, or equivalently $E$ must be
less than $B$.  Generally speaking, eq.~(\ref{eq:Motion:general}) can
have solutions where $|\Ud|>c$. The surface, where $|\Ud|$ reaches
$c$, is commonly referred as the \emph{light surface}. Beyond the
light surface, where $|\Ud|>c$, the force-free approximation can not
be applied. FFE is not self-consistent, because particle dynamics is
ignored. Hence, each solution of eq.~(\ref{eq:Motion:EB}) should be
always checked for applicability of the force free approximation.

In the axisymmetric stationary case considered here magnetic field in
cylindrical coordinates ($\varpi,\phi,Z$) can be written as
\begin{equation}
  \label{eq:B}
  \B = \frac{ \vec{\nabla} \Psi \times \vec{e_\phi} }{ \varpi } +
  \frac{4\pi}{c} \frac{I}{\varpi} \vec{e_\phi}
  \,,
\end{equation}
where $\vec{e_\phi}$ is the unit azimuthal, toroidal vector. In
components
\begin{equation}
  \label{eq:B_components}
  (B_\varpi, B_\phi, B_Z) = 
  (-\frac{1}{\varpi}\pd_Z\Psi,\: 
  \frac{4\pi}{c}\frac{I}{\varpi},\: 
  \frac{1}{\varpi}\pd_\varpi\Psi)
  \, .
\end{equation}
Scalar function $\Psi$ is related to the magnetic flux
$\Phi_\mathrm{mag}$ trough the circle with the centre at the point (0,
$Z$) and radius $\varpi$ by $\Phi_\mathrm{mag}=2\pi\Psi(\varpi,Z)$.
So, lines of constant $\Psi$ coincides with magnetic field lines.  It
could be easily verified, that in force-free case the scalar function
$I(\varpi,Z)$ is constant along magnetic field lines, i.e
\begin{equation}
  \label{eq:I(Psi)}
  I \equiv I(\Psi)
  \, .
\end{equation}
$I$ is related to the total current $J$ \emph{outflowing} trough the
above mentioned circle by $J=2\pi{}I(\varpi,Z)$.

In the quasi-stationary case the time derivative of B takes the form
\citep[see][]{Mestel73}
\begin{equation}
  \label{eq:dB/dt}
  \frac{\pd \B}{\pd t} = 
  \nabla \x ([\vec{\Omega} \x \vec{r}] \x \B)
  \, .
\end{equation}
Substituting this into the Faraday's law
\begin{equation}
  \label{eq:Maxwell_Farad}  
  \nabla \x \E = -\frac1{c} \pd_t \B 
  \, ,
\end{equation}
we get for the electric field
\begin{equation}
  \label{eq:E:general}
  \E = - \frac{\vec{\Omega} \x \vec{r}}{c} \x \B - \nabla V =
  - \frac{\Omega}{c} \nabla\Psi - \nabla V
  \, ,
\end{equation}
where $V$ is the non-corotational (see below) part of electric
potential. The first term in (\ref{eq:E:general}) is poloidal and only
the second term could make a contribution to the toroidal component.
In axisymmetric case $\pd_\phi{}V=0$ and, hence, \E{} is poloidal.  In
the force-free case $\E\perp\B$, from this follows that
$\E\cdot(\nabla\Psi\x\vec{e_\phi})=0$. Consequently,
$\E\propto\nabla\Psi$ and we can write
\begin{equation}
  \label{eq:E}
  \E = - \frac{\OmF}{c} \nabla \Psi
  \, ,
\end{equation}
or in components
\begin{equation}
  \label{eq:E_components}
  (E_\varpi,E_\phi,E_Z) = 
  (-\frac{\OmF}{c}\pd_\varpi\Psi,\:  
  0,\: 
  -\frac{\OmF}{c}\pd_Z\Psi)
  \, .
\end{equation}
Substituting this expression together with eq.~(\ref{eq:B}) into the
formula for the drift velocity~(\ref{eq:Udrift:general}), we get
\begin{equation}
  \label{eq:Udrift:Omega_B}
  \Ud = \OmF \varpi \Uvec{\phi} - 
  \frac{4\pi}{c}\frac{I \OmF}{B^2}\; \B \equiv
  \vec{\OmF} \x \vec{r} -\kappa \B
  \, .
\end{equation}
So, the particle motions is composed from rotation with the angular
velocity \OmF{} and gliding along magnetic field lines. Hence, \OmF{}
is the angular velocity of magnetic field lines rotation. By
substitution of eq.(\ref{eq:E}) into the stationary Faraday's law one find
$\nabla\OmF\x\nabla\Psi=0$. This implies that \OmF{} is constant along
magnetic field lines:
\begin{equation}
  \label{eq:Om(Psi)}
  \OmF \equiv \OmF(\Psi)
  \,.
\end{equation} 
Equation~(\ref{eq:Om(Psi)}) is the well known Ferraro isorotation law.

Finally, substituting \E{} and \B{} from eqs.~(\ref{eq:B}),
(\ref{eq:E}) into equation~(\ref{eq:Motion:EB}) we get
\begin{eqnarray}
  \left( 1-\frac{\OmF^2 \varpi^2}{c^2} \right) \nabla^2\Psi -
  \frac{2}{\varpi} \pd_\varpi \Psi +\qquad\qquad\qquad
  &&\nonumber\\
  \qquad{}+\left( \frac{4\pi}{c} \right)^2 I \frac{d I}{d \Psi}  
  -\frac{\varpi^2}{c^2} \OmF \frac{d \OmF}{d \Psi} 
  \left( \nabla \Psi \right)^2 & = & 0
  \label{eq:PulsarEq_Full}
\end{eqnarray}
This is Grad-Shafranov equation for the poloidal magnetic field, the
so-called pulsar equation, derived by
\citet{Michel73,ScharlemannWagoner73,Okamoto74}. This scalar PDE is of
elliptical type. It is the poloidal part of the vector
equation~(\ref{eq:Motion:EB}). The toroidal part of
eq.~(\ref{eq:Motion:EB}) is simply the relation (\ref{eq:I(Psi)}).
Pulsar equation has two integrals of motion -- $I$ and \OmF{}. If we
know them, we can solve this equation for function $\Psi$ and
determine the poloidal magnetic field. Electric and magnetic fields
and all other parameters of the force-free magnetosphere can be found,
because they are connected to $\Psi$, $I$ and \OmF{} by algebraic
relations.  In the frame of FFE $I$ and \OmF{} are \emph{free
  parameters}. They could be determined self-consistently in the full
MHD, if also electromagnetic cascades, setting boundary conditions,
are taken into account \citep[see][]{BeskinBook}. Nevertheless one can
get useful results in the force-free approximation.
Equation~(\ref{eq:PulsarEq_Full}) has one singular surface, the
so-called \emph{light cylinder} (LC), where $\varpi =
c/\OmF(\Psi(\varpi,Z))$.  As it will be shown in the next subsection
the difference between \OmF{} and $\Omega$ is small and the singular
surface has shape close to a cylinder with the radius of
$\RLC=c/\Omega$.

We normalise variables $\varpi$ and $Z$ to \RLC{} and introduce new
dimensionless coordinates $x\equiv \varpi/\RLC$ and $z\equiv Z/\RLC$.
We will consider the case of dipolar magnetic field on the NS. So,
near the star the magnetic field is given by
\begin{equation}
  \label{eq:PsiDipol}
  \Psi = \mu \frac{\varpi^2}{(\varpi^2+Z^2)^{3/2}} \equiv
  \Psi_0 \frac{x^2}{(x^2+z^2)^{3/2}}
  \,,
\end{equation}
where $\mu=B_0\RNS^3/2$ is the magnetic moment of the NS and
$\Psi_0\equiv\mu/\RLC$. We normalise $\Psi$ to $\Psi_0$ and introduce
dimensionless function $\psi\equiv\Psi/\Psi_0$. Instead of poloidal
current function $I$ we introduce dimensionless function
$S\equiv(4\pi/c)(\RLC/\Psi_0)I$.  Angular velocity of magnetic field
line rotation is normalised to the angular velocity of the NS by the
relation $\OmF(x,z)\equiv\beta(x,z)\,\Omega$.  For these dimensionless
functions the pulsar equation~(\ref{eq:PulsarEq_Full}) takes the form
\begin{eqnarray}
  \label{eq:PsrEq_Full:dimensionless}
  (\beta^2 x^2-1)(\pd_{xx}\psi + \pd_{zz}\psi) +
  \frac{\beta^2 x^2+1}{x} \pd_x \psi -
  && \nonumber\\
  \quad\quad\quad
  - S \frac{d S}{d \psi}
  + x^2 \beta \frac{d \beta}{d \psi} \left( \nabla \psi \right)^2
  & = & 0
  \,.
\end{eqnarray}
At the light cylinder the coefficient by second derivatives goes to
zero and the pulsar equation has the form
\begin{equation}
  \label{eq:CondAtLC_Full:dimensionless}
  2\beta \, \pd_x \psi = 
  S \frac{d S}{d \psi}
  - \frac{1}{\beta} \frac{d \beta}{d \psi} \left( \nabla \psi \right)^2
  \,.
\end{equation}
Let us now discuss properties of functions \OmF{} and $S$.

\subsection{\OmF}
\label{sec:omf}

From relations~(\ref{eq:Om(Psi)}) it follows that $V$ is constant along
a magnetic field line. Hence, we could rewrite eq.~(\ref{eq:E:general})
in the following form
\begin{equation}
  \E =  - \frac1{c}\left( \Omega + c \frac{\pd V}{\pd \Psi} \right)
  \nabla\Psi
  \,.
\end{equation}
Comparing this expression with eq.~(\ref{eq:E}) we get
\begin{equation}
  \label{eq:OmF}
  \OmF = \Omega + c \frac{\pd V}{\pd \Psi}
  \,.
\end{equation}
If there were no potential difference between different magnetic field
lines and between them and the surface of the pulsar, \OmF{} were
equal to $\Omega$.  But, independently of NS surface properties, a
potential difference along open magnetic field lines will be always
build in polar cap region of pulsar
\citep{Ruderman/Sutherland75,Arons/Scharlemann/78,Muslimov/Tsygan92}.
This lead to formation of a particle acceleration zone, where
force-free approximation is not valid and charged particles are
accelerated by the longitudinal electric field. Electron-positron
pairs produced in the strong magnetic field of pulsar by photons,
emitted by accelerated particles, screen the accelerating field, and
as pair-production rate grows very rapidly with the distance,
acceleration zone terminates in a rather thin layer called
pair-formation front (PFF). Above PFF accelerating field is screened
and FFE can be applied.  The size of the acceleration zone is small
compared to the overall size of the magnetosphere, its height varies
from $\sim 100$~m for young pulsars in model with no particle escape
from the NS surface \citep{Ruderman/Sutherland75} to 1-2 stellar radii
in models, where particles freely escape the star surface
\citep{Arons/Scharlemann/78,Muslimov/Tsygan92}.  Geometrically this
small region could be neglected in the modelling of the global
magnetospheric structure. The potential difference between NS surface
and magnetic field lines should be taken into account by boundary
conditions on $V$, which can be reformulated as boundary conditions on
\OmF{}. Potential difference along a magnetic field line in the
acceleration zone is determined by the position of PFF, which depends
on local geometry of magnetic field, close to the NS surface, and
kinetic processes in the electron-positron cascade.

By the order of magnitude the relative difference of rotation
velocities of plasma and NS can be estimated as
\begin{equation}
  \label{eq:dOm_Om}
  \begin{array}{l}
    \displaystyle
    \frac{\delta \Omega}{\Omega} \equiv
    \frac{\Omega-\OmF}{\Omega}   \simeq 
    \frac{P}{2\pi}\frac{v_\mathrm{rot}}{\RPC}  \\
    \displaystyle
    \quad\quad\quad\quad\quad\quad
    \simeq 2.28\x10^{-11} 
    \left( \frac{B_0}{10^{12}\,\mathrm{G}} \right)^{-1} P^2 \D V
    \,,
  \end{array}
\end{equation}
where $\D{}V$ is potential difference between NS surface and PFF (in
esu units), $\RPC \simeq \sqrt{\RNS^3\Omega/c}$ is the size of the
polar cap; $v_\mathrm{rot}=c\D{}V/(B_0\RPC)$ is the linear velocity of
plasma rotation relative to the NS surface in the acceleration zone --
see eq.~(31) in \citet{Ruderman/Sutherland75}.

In the model with no particle escape from the NS surface the potential
difference is given by \citep[equation~(23)]{Ruderman/Sutherland75}
\begin{equation}
  \label{eq:dV_RS}
  \D V\simeq 5.24\x 10^{9} P^{-1/7} 
  \left( \frac{\rho_\mathrm{c}}{10^6 \mathrm{cm}} \right)^{4/7}
  \left( \frac{B_0}{10^{12}\,\mathrm{G}} \right)^{-1/7} 
  ,
\end{equation}
where $\rho_\mathrm{c}$ is the curvature radius of magnetic field
lines. The potential difference is measured in esu units. Substituting
these expression into eq.~(\ref{eq:dOm_Om}) we get
\begin{equation}
  \label{eq:dOm_Om__RS}
  \frac{\delta \Omega}{\Omega}   \simeq
  0.1\; P^{13/7}
  \left( \frac{\rho_\mathrm{c}}{10^6 \mathrm{cm}} \right)^{4/7}
  \left( \frac{B_0}{10^{12}\,\mathrm{G}} \right)^{-8/7}
  .
\end{equation}
We see, that for relatively young pulsars, with periods $P\la0.3$~s,
this ratio is very small, $\sim1$ per cent. Even if the field line
curvature radius is of the order of $\sim 10^8$~cm, typical for dipole
magnetic field, for $P\la0.1$~s this ratio is $\sim2$ per cents.

For the model where particles freely escape the NS surface we use
estimations from \citet{Hibschman/Arons01}.  The potential difference
in the acceleration zone \citep[eqs.~(17) and (18)]{Hibschman/Arons01}
\begin{eqnarray}
  \label{eq:dV_Ah}
  \D V^{h>\RPC} & \simeq & 
  9.87\x 10^9 P^{-2}
  \left( \frac{B_0}{10^{12}\,\mathrm{G}} \right)
  h
  \,,\\
  \D V^{h<\RPC} & \simeq & 
  1.11\x 10^{12} P^{-3/2}
  \left( \frac{B_0}{10^{12}\,\mathrm{G}} \right)
  h^2
  \,.
\end{eqnarray}
Here $h$ is the height of PFF above the NS surface in units of \RNS{}.
The above estimations for accelerating potential are for the cases
when when $h>\RPC$ and $h<\RPC$ correspondingly. The potential
differences are in esu units. The heights of PFF position due
to photons emitted by non-resonant inverse Compton scattering (NIC),
curvature radiation (CR) and resonant inverse Compton scattering (RIC)
of accelerated particles are given by
\begin{eqnarray}
  \label{eq:h_A__hNIC}
  h_\mathrm{NIC} & \simeq & 
  0.40\, P
  \left( \frac{B_0}{10^{12}\,\mathrm{G}} \right)^{-1}
  T_6^{-2}
  f_\rho \\
  \label{eq:h_A__cNIC}
  h^\mathrm{c}_\mathrm{NIC} & \simeq &
  0.12\, P^{1/4}
  \left( \frac{B_0}{10^{12}\,\mathrm{G}} \right)^{-1/2}
  T_6^{-1}
  f_\rho^{1/2} \\  
  \label{eq:h_A__CR}
  h_\mathrm{CR} & \simeq &
  0.68\, P^{19/12}
  \left( \frac{B_0}{10^{12}\,\mathrm{G}} \right)^{-5/6}
  f_\rho^{1/2}\\  
  \label{eq:h_A__RIC}
  h_\mathrm{RIC} & \simeq &
  12\,
  \left( \frac{B_0}{10^{12}\,\mathrm{G}} \right)^{-7/3}
  T_6^{-2/3}
  f_\rho
  \,,
\end{eqnarray}
see \citet{Hibschman/Arons01}, eqs.~(34), (32), (42) and (37)
correspondingly.  Label ``c'' correspond to the model where the NS
surface is colder than the polar cap of the pulsar, heated by the
return current. $T_6$ is the temperature of the polar cap in units of
$10^6$~K. The radius of curvature of magnetic field lines is factor
$f_\rho$ times the radius of curvature of a dipole field, i.e.
$f_\rho\equiv \rho_c/\rho_c^{dip} = P^{-1/2}\,\rho_c/(9.2\x10^7
\mathrm{cm})$.

According to \citet{Hibschman/Arons01}, in most pulsar the PFF height
is set by non-resonant ICS photons.  In high voltage pulsar, ones with
the shortest periods -- millisecond and youngest pulsar with $P\la0.3$~s,
the PFF is set by curvature photons.  In both of these cases the
resulting height of the PFF is larger, than the size of the polar cap,
$h>\RPC$.  Resonant ICS is important only for high field pulsars, with
$B\ga1.2\x10^{13}$~G, in this case $h\ll\RPC$.  Taking this into
account we get
\begin{eqnarray}
  \label{eq:dOmOm_A__hNIC}
  \left( 
    \frac{\delta \Omega}{\Omega}
  \right)_\mathrm{NIC} 
  & \simeq & 
  0.09\, P
  f_\rho\, 
  T_6^{-2}
  \left( \frac{B_0}{10^{12}\,\mathrm{G}} \right)^{-1}
  \\
  \label{eq:dOmOm_A__cNIC}
  \left( 
    \frac{\delta \Omega}{\Omega}
  \right)^\mathrm{c}_\mathrm{NIC} 
  & \simeq & 
  0.027\, P^{1/4}
  f_\rho^{1/2}\, 
  T_6^{-1}
  \left( \frac{B_0}{10^{12}\,\mathrm{G}} \right)^{-1/2}
  \\  
  \label{eq:dOmOm_A__CR}
  \left( 
    \frac{\delta \Omega}{\Omega}
  \right)_\mathrm{CR\ } 
  & \simeq &
  0.023\, 
  \left(\frac{P}{0.3\, \mathrm{s}} \right)^{19/12}
  f_\rho^{1/2} 
  \left( \frac{B_0}{10^{12}\,\mathrm{G}} \right)^{-5/6}
  \\
  \label{eq:dOmOm_A__RIC}
  \left( 
    \frac{\delta \Omega}{\Omega}
  \right)_\mathrm{RIC} 
  & \simeq &  
  0.034\, P^{1/2}
  f_\rho^2\, 
  T_6^{-4/3}
  \left( \frac{B_0}{1.2\x10^{13}\,\mathrm{G}} \right)^{-14/3}
  \,.
\end{eqnarray}
The temperature of the polar cap $T$ due to the heating by return
particles is of the order of $10^6$~K. If the NS temperature is higher
than this value, than formula~(\ref{eq:dOmOm_A__hNIC}) should be
applied, in the opposite case -- formula~(\ref{eq:dOmOm_A__cNIC}).
The temperature of the NS surface depend of neutron star cooling
model, and for rather young pulsar it should be higher than $10^6$~K.
So, formula~(\ref{eq:dOmOm_A__hNIC}) is applicable for young, hot
pulsar, where it gives for $\delta\Omega/\Omega\simeq0.01$.  Hence, in
the model with free particle escape, the ratio $\delta \Omega/\Omega$
is of the order of few per cents for the majority of pulsars.

We see, that $1-\beta$ is of order of few per cents for most pulsars
in the model with free particle escape and for young pulsars in the
model with no particles escape.  We restrict ourself considering only
such pulsars, where $1-\beta$ is small. Then the last term in pulsar
equation~(\ref{eq:PsrEq}) is small in comparison with other terms and
could be neglected.  In the rest of the paper we assume
\begin{equation}
  \label{eq:OmF=Om}
  \OmF\equiv\Omega
  \,.
\end{equation}
This assumption simplifies the pulsar
equation~(\ref{eq:PsrEq_Full:dimensionless}), which now has the form
\begin{equation}
  \label{eq:PsrEq}
  (x^2-1)(\pd_{xx}\psi + \pd_{zz}\psi) + 
  \frac{x^2+1}{x} \pd_x \psi - \SSp  =  0
  \,,  
\end{equation}
where $S^\prime \equiv d S/d \psi$.  Nonlinearity in this equation is
now present only in the term with the poloidal current function $S$.

%
\begin{figure}
  \includegraphics[clip,width=\columnwidth]{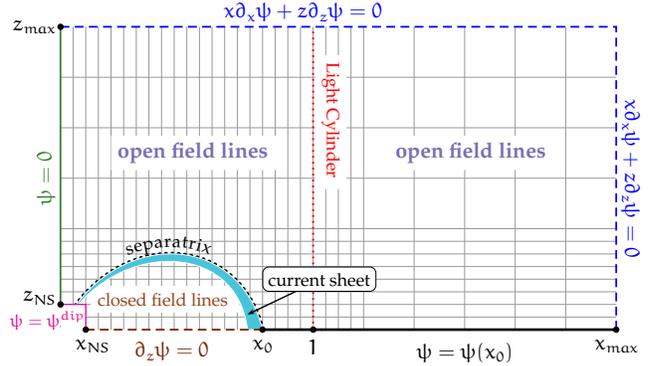}
  \caption{Calculation domain and imposed boundary conditions. See
    text for explanation.}
  \label{fig:CalcDomain}
\end{figure}

\subsection{Poloidal current $S$}
\label{sec:poloidal-current-s}

In contrast to \OmF{}, being set by kinetic processes in the polar
cap, $S$ depends on the global structure of the magnetosphere.  Both
inside and outside the light cylinder the pulsar
equation~(\ref{eq:PsrEq}) is a regular non-linear PDE of elliptic
type.  At the light cylinder this equation under
assumption~(\ref{eq:OmF=Om}) has the form
\begin{equation}
  \label{eq:CondAtLC}
  \pd_x \psi = \frac1{2} \SSp
  \,.
\end{equation}
If function $S$ is known, condition~(\ref{eq:CondAtLC}) can be
considered as a Neumann type boundary condition at the light cylinder.
If boundary conditions are set both inside and outside the LC, the
equation should have an unique solution in both regions. Generally
speaking, for arbitrary function $S$ solutions of the pulsar equation
inside and outside the LC will not match,
\begin{equation}
  \label{eq:DiscontinuityAtLC}
  \lim_{x\rightarrow 1^-} \psi \neq \lim_{x\rightarrow 1^+}\psi
  \,.
\end{equation}
So, a smooth%
\footnote{if solution is continuous its smoothness follows from
  eq.~(\ref{eq:CondAtLC}), because \SSp{} is the same at both sides of
  the LC.}
solution is possible only for a specific function $S$ and the problem
of finding a solution of the pulsar equations becomes an eigenvalue
problem for the function $S$.

The position of the light cylinder is not known
a priory. For \OmF{} different from $\Omega$ it has a rather
complicated form and, even if $\OmF(\Psi)$ as a function of $\Psi$ is
given by a model of the polar cap cascade, the position of the LC as a
function of $x$ and $z$ has to be found self-consistently together
with the solution of the pulsar equation. However, as it was stressed
above, for most pulsar the deviation of the LC from a cylinder with
the radius $c/\Omega$ is of the order of few per cents or less. Hence,
a solution of equation (\ref{eq:PsrEq}) should give a very good
approximation to the real magnetosphere of aligned rotator.

The other open question regarding the poloidal current term in the
pulsar equation is the topology of the magnetosphere. In works of the
Lebedev Physical Institute group \citep[see
e.g.][]{GurevichBeskinIstomin_Book,Beskin/Malyshkin98} a geometry with
X null point have been assumed, hereafter X-configuration, see
Fig.~\ref{fig:Topology-X}(a). In that case the pulsar equation should
be solved in 3 different domains, separated by the current sheets. The
positions of the point $A$(and $A^\prime$) is a free parameter of such
model. Setting the positions of these points and the point $x_0$ one
fixes the boundaries and gets a well posed, although complicated,
problem.  Such topology of the aligned rotator magnetosphere was
criticised by \citet{Lyubar90}, because the only source of the
magnetic field in the magnetosphere is the pulsar itself, and in this
case it is not clear what would be the source of the magnetic field in
the outer domain.  The most frequently considered topology of the
aligned rotator magnetosphere implies an Y-like null point, hereafter
Y-configuration, see Fig.~\ref{fig:Topology-Y}.  In this case the only
available free parameter in the model is the position of the null
point $x_0$. Fixing position of this point we fix the whole geometry
of the magnetosphere.  So, we have an elliptic equation with boundary
conditions set at all boundaries of the closed domains with known
positions of the boundaries. We wish to emphasise here, that the
choice of the magnetosphere's topology is an additional assumption in
the frame of stationary problem. In the following we investigate in
details the force-free magnetosphere of aligned rotator assuming
topology with an Y-like neutral point.

\section{Numerical Model}
\label{sec:numerical-model}

We solve equation (\ref{eq:PsrEq}) in a rectangular domain, see
Fig.~\ref{fig:CalcDomain}. The boundary conditions 
are the following. On the rotation axis (z-axis)
\begin{equation}
  \label{eq:bcond_zaxis}
  \psi(0,z) = 0, \qquad \ZNS<z \le \ZMax
  \,.
\end{equation}
At the equatorial plane, in the closed field line zone
\begin{equation}
  \label{eq:bcond_xclosed}
  \pd_z \psi(x,0)= 0, \qquad \XNS<x<\XL
  \,,
\end{equation}
following from the symmetry of the system. In the open field line
domain
\begin{equation}
  \label{eq:bcond_xopen}
  \psi(x,0)=\psi(x_0,0), \qquad \XL<x \le \XMax
  \,,
\end{equation}
i.e. the separatrix lies in the equatorial plane. Close to the NS the
magnetic field is assumed to be dipolar, so for
$x=\XNS,0\le{}z\le\ZNS$ and $0\le{}x\le\XNS,z=\ZNS$
\begin{equation}
  \label{eq:bcond_center}
  \psi(x,z)=\psi^\mathrm{dip}(x,z) \equiv
  \frac{x^2}{(x^2+z^2)^{3/2}}
  \,.
\end{equation}
Magnetic surfaces should become radial at large distance from the NS,
see \citet{Ingraham73}. On the other hand, in calculations of
\citet{CKF}, where pulsar equation was solved in the unbounded domain,
with boundary conditions at infinity implying finiteness of the total
magnetic flux, magnetic surfaces became nearly radial already at
several sizes of the light cylinder. Rather different outer boundary
conditions, with finite magnetic flux inside the light cylinder at
infinity, have been used by \cite{Lovelace90}.  However,
time-dependent simulations of \citet{Komissarov05,Spitkovsky05_poland}
and \citet{McKinney:NS:06} provide strong evidence for correctness of
outer boundary conditions when magnetic surfaces at large distances
from the NS are radial.  So, at the outer boundaries of the
calculation domain for $0<x\le\XMax,z=\ZMax$ and $x=\XMax,0<z\le\ZMax$
\begin{equation}
  \label{eq:bcond_outer}
  x\,\pd_x\psi + z\,\pd_z\psi = 0.
\end{equation}

At the light cylinder two conditions should be satisfied: (i) the
solution should be continuous,
\begin{equation}
  \label{eq:LC_cont}
  \psi(x\rightarrow 1^-,z)=\psi(x\rightarrow 1^+,z)
  \,,
\end{equation}
and (ii) the condition (\ref{eq:CondAtLC}). These conditions together
provide smooth transition through the LC. Following \citet{Goodwin/04}
we expand function $\psi$ at the LC in Taylor series over $x$ implying
continuity condition~(\ref{eq:LC_cont}). By substituting the resulting
expansion into the pulsar equation~(\ref{eq:PsrEq}) and retaining the
terms up to the second order we get the following approximation to the
pulsar equation at the LC
\begin{equation}
  \label{eq:PsrEq_LC}
  4\; \pd_{xx}\psi(1,z) + 2\; \pd_{zz}\psi(1,z) = 
  \pd_{x}\left[ \SSp(1,z) \right]
  \,.
\end{equation}
This equation is nothing more than a reformulation of the smoothness
conditions~(\ref{eq:LC_cont}), (\ref{eq:CondAtLC}) valid for the first
and second order terms in Taylor series expansion of $\psi$.  As the
numerical scheme we have used is of the second order, this
approximation, as well as its discretization, has the same accuracy as
the discretized equation in the rest of the numerical domain. In
course of relaxation procedure we are trying to satisfy the conditions
(\ref{eq:LC_cont}), (\ref{eq:CondAtLC}), i.e.  we solve
equation~(\ref{eq:PsrEq_LC}) at the LC instead of the original
equation~(\ref{eq:PsrEq}), which is singular there.
Equation~(\ref{eq:CondAtLC}) is used for determination of the poloidal
current term $\SSp(\psi)$ along the open field lines.

In the closed field lines zone, $\psi>\PsiL \equiv\psi(\XL,0)$, there
is no poloidal current, so $\SSp \equiv 0$. The return current needed
to keep the system charge neutral flows along the separatrix.  In the
open field line domain by setting the boundary
condition~(\ref{eq:bcond_xopen}) the presence of an infinite thin
current sheet is already incorporated into the solution procedure.
However, when the separatrix goes above the equatorial plane we have
to model the current sheet. We assume that the return current is
flowing along the field lines corresponding to the magnetic surfaces
$[\PsiL,\PsiL+d\psi]$. The total return current flowing in this region
is calculated by integration of the term \SSp{}:
\begin{equation}
  \label{eq:Spoloidal}
  S_\mathrm{return} = \sqrt{ 2 \int_{0}^{\PsiL} \SSp \, d\psi}
  \;.
\end{equation}
We model the poloidal current density distribution over $\psi$ in the
current sheet $\PsiL \le \psi \le \PsiL+d\psi$ by an even order
polynomial function going to zero at the boundaries of the current
sheet
\begin{equation}
  \label{eq:S_CurrentSheet}
  S^\prime(\psi)= 
  A 
  \left[
    \left(\psi-\left(\PsiL+\frac{d\psi}{2}\right)\right)^{2k} -
    \left(\frac{d\psi}{2}\right)^{2k}
  \right]
  \,,
\end{equation}
where constant $A$ is determined from the requirement
$\int_{0}^{\PsiL+d\psi} S(\psi) \, d\psi = 0$ and $k$ is an integer
constant.  The pulsar equation is then solved in the whole domain
including the current sheet.  Although the current sheet cannot be
considered as a force-free domain, but doing so we calculate correct
the influence of the current sheet on to the global magnetospheric
structure, though the obtained values of the physical parameters
\emph{inside} the current sheet are fake.

We developed a multigrid numerical scheme for solution of
equations~(\ref{eq:PsrEq}) and (\ref{eq:PsrEq_LC}). These equations
have been discretized using the 5-point Gauss-Seidel rule. The
coarsest numerical grid was constructed in the way, that the light
cylinder is at cell boundaries. Each subgrid was obtained by halving
of the previous grid. Cell sizes in the region $x<1,z<1$ are smaller in
order to accurate calculate the current along the separatrix.  We used
FAS scheme with V-type cycles \citep[see][]{TrottenbergBook}.  The
Gauss-Seidel scheme was used as both smoother and solver at the
coarsest level.  At each iteration step both in the solver and
smoother the new value of the poloidal current term $\SSp(1,z)$ was
calculated from the relation~(\ref{eq:CondAtLC}) at each point of the
LC. Then a piece-polynomial interpolation of \SSp{} in the interval
$(0,\PsiL)$ was constructed and the return current distribution was
calculated according to the formulae~(\ref{eq:Spoloidal}) and
(\ref{eq:S_CurrentSheet}). Then for each point $(x,z)$ in the
calculation domain the current term was calculated as
$\SSp(x,z)=\SSp(\psi(x,z))$, and the new iteration was started.  So,
we solved the pulsar equation in the whole domain avoiding a very time
consuming matching of the solutions inside and outside the light
cylinder as it was done by \citet{CKF,Contopoulos05} and
\citet{Gruzinov:PSR}, though in \citet{Contopoulos05} this matching
procedure have been accelerated. As a starting configuration a dipolar
magnetic field everywhere was used.  We did not encounter any problems
with the convergence of the scheme for any value of \XL{}, but for
\XL{} very close to 1 the convergence rate becomes essentially slower.
Typical number of points along each directions we used in the
calculations was $3000-6000$.

We performed calculations for different values of numerical parameters
in order to proof the independence of the results on the domain sizes
$(\XMax,\ZMax)$, the ``NS size'' $(\XNS,\ZNS)$, the width of the
current sheet $d\psi$ and the form of the current distribution
(parameter $k$), as well as on the iteration procedure stopping
criteria and number of points along both directions. Changes in
convergence criteria and decreasing of the cell size from ones used in
the most of our calculations did not produce relative changes in
solutions greater that $10^{-4}$. In Table~\ref{tab:SolProp} values of
\PsiL{} and energy losses of aligned rotator $W$ (see next section),
obtained in computations with different values of listed numerical
parameters, are shown for $\XL=0.7$ and \XL{} approaching the light
cylinder. One can see, that with an accuracy of the order of few per
cents obtained solutions are independent on particular values of the
numerical parameters. Solutions with other \XL`s have similar
behaviour.

\begin{table}
  \caption{Properties of obtained solution with \XL=0.7 and \XL{} approaching the
    LC for different values of numerical parameters} 
  \begin{tabular}{cccccc}
    \hline
    \multicolumn{4}{|c|}{numerical parameters:} & \multicolumn{2}{|c|}{results:}  \\
    $d\psi$ & $2k$ & $(x_{max},z_{max})$ & $(x_{NS}, z_{NS})$ & \PsiL
    & $W$ \\
    \hline
    \multicolumn{6}{l}{\XL=0.7}   \\
    \hline
    0.03&    2 &              (8,7) & (0.0667, 0.056)    &   1.717 & 1.864 \\
    0.03&    4 &              (8,7) & (0.0667, 0.056)    &   1.712 & 1.853 \\
    0.015&   2 &              (8,7) & (0.0667, 0.056)    &   1.697 & 1.821 \\
    0.03&    2 &              (8,7) & (0.0333, 0.028)    &   1.720 & 1.870 \\
    0.03&    2 &            (16,14) & (0.0667, 0.056)    &   1.717 & 1.864  \\
    \hline

    \multicolumn{6}{l}{\XL=0.99} \\
    \hline
    0.08&    2 &            (16,14) & (0.06, 0.06)    &   1.255 & 0.977  \\
    \hline
    \multicolumn{6}{l}{\XL=0.99231} \\
    \hline
    0.04&    2 &            (5,5) & (0.0462, 0.0525)    &   1.230 & 0.939  \\
    \hline
    \end{tabular}
  \label{tab:SolProp}
\end{table}

\section{Results of calculations}
\label{sec:results-calculations}

%
\begin{figure*}
  \begin{center}
    \includegraphics[clip,totalheight=5.1cm]{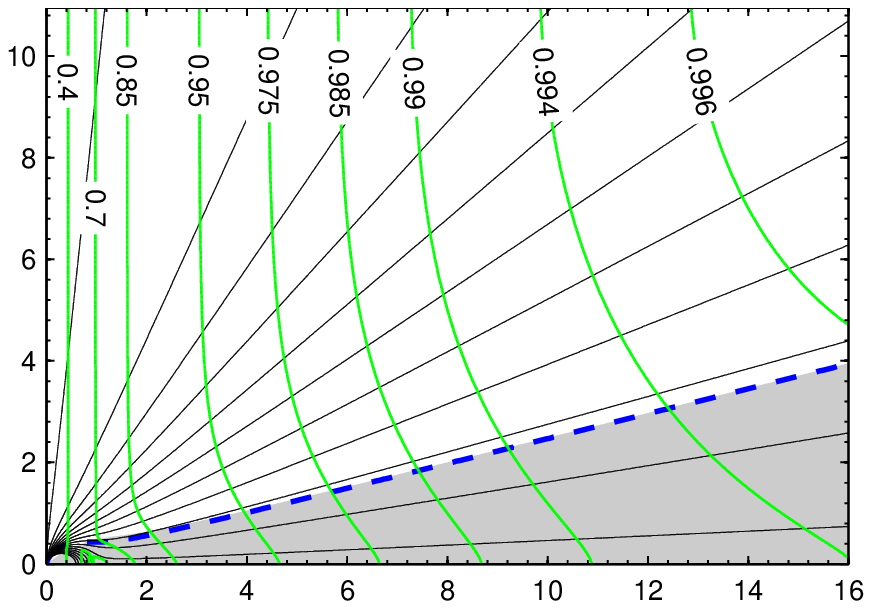}
    \includegraphics[clip,totalheight=5.1cm]{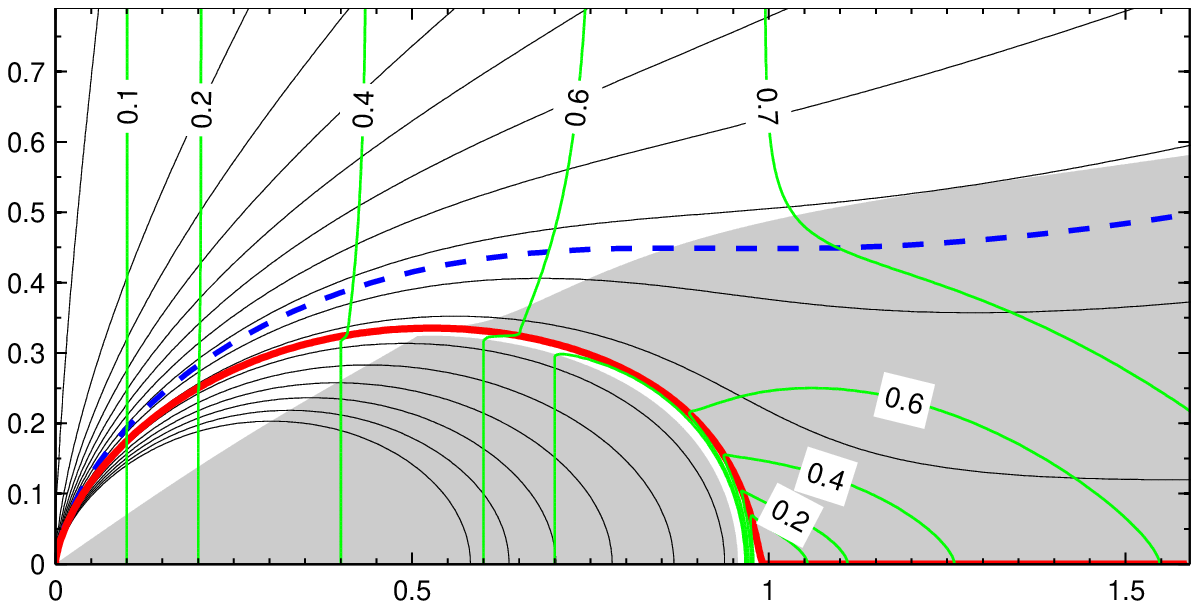}
    \includegraphics[clip,totalheight=5.1cm]{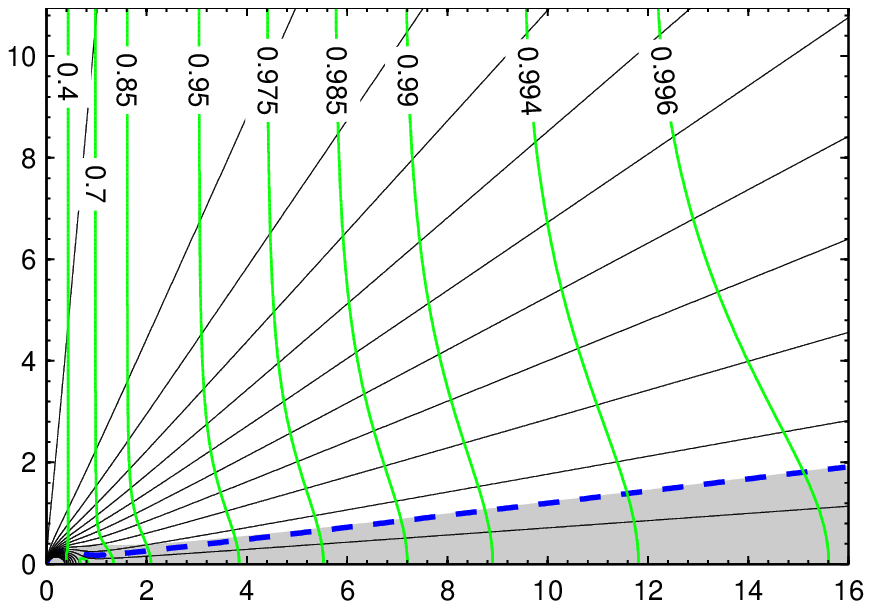}
    \includegraphics[clip,totalheight=5.1cm]{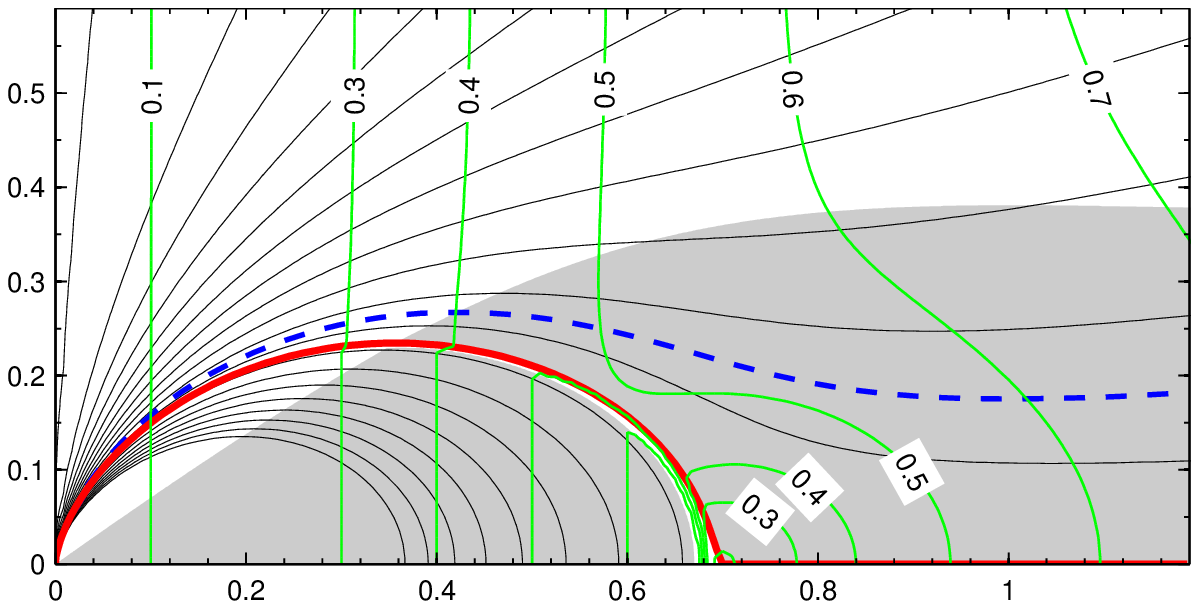}
    \includegraphics[clip,totalheight=5.1cm]{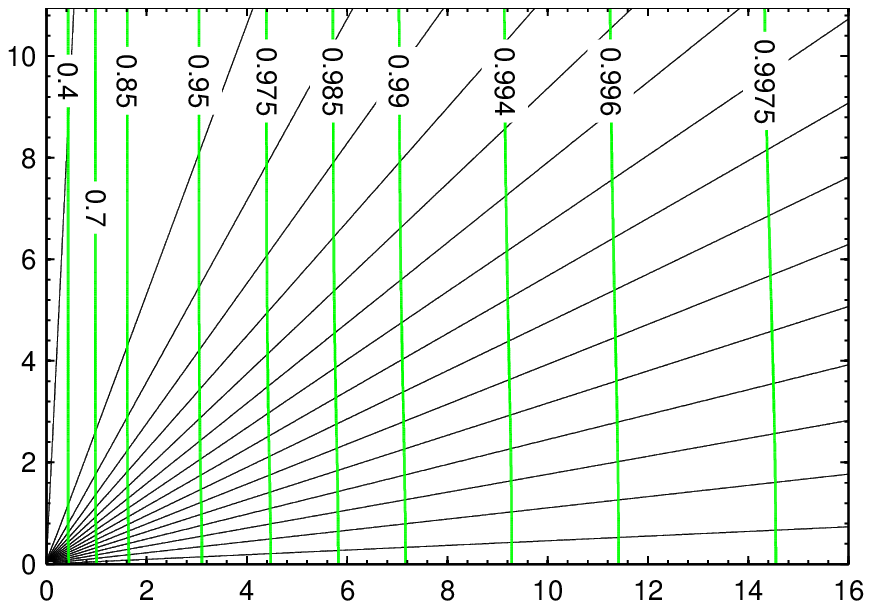}
    \includegraphics[clip,totalheight=5.1cm]{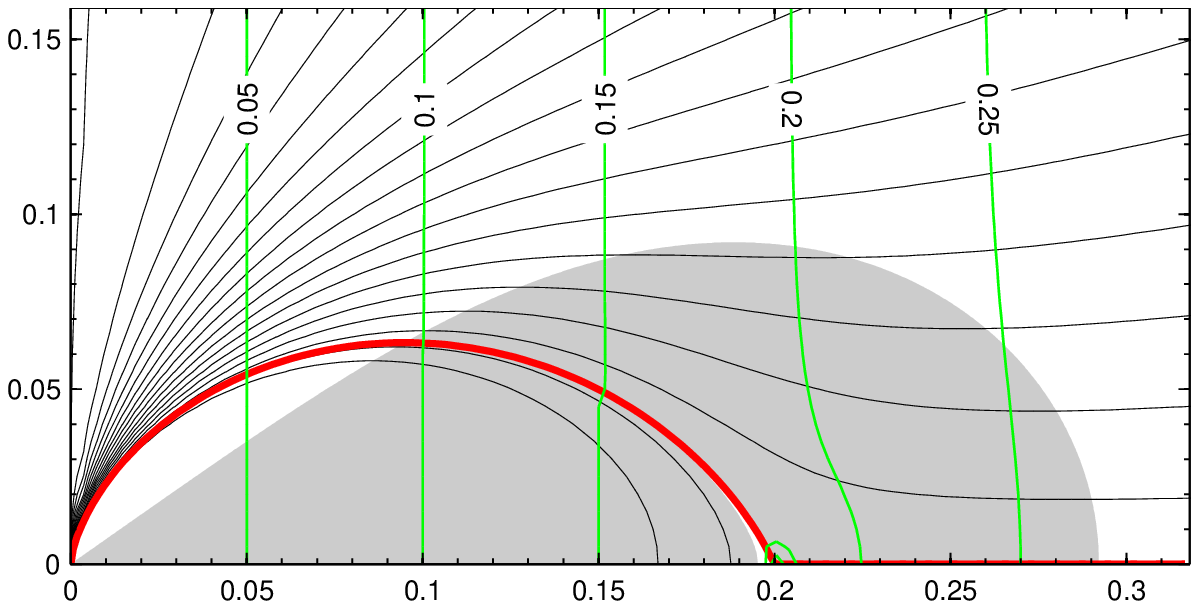}
  \end{center}
  \caption{Global structure of the magnetosphere for $x_0=0.992$ --
    top figures, $x_0=0.7$ -- middle figures, $x_0=0.2$ -- bottom
    figures.  Magnetic flux surfaces are shown by thin solid lines,
    the labelled vertical lines are contours of the drift velocity,
    grey area is the domain where the GJ charge density is positive.
    Dashed line separates regions with direct (above the line) and
    return (below the line) volume currents. The separatrix is shown
    by the thick solid line. On the left figures almost the whole
    calculation domain is shown, on the right figures -- the central
    part of the calculation domain. Distances along $x$-axis
    (horizontal) and $z$-axis (vertical) are measured in LC radius
    \RLC.}
  \label{fig:Main}
\end{figure*}

Our choice of the boundary conditions at the NS,
equation~(\ref{eq:bcond_center}), corresponds to the case when the
dipole magnetic moment of the star $\vec{\mu}$ is parallel to the
angular velocity vector \vec{\Omega}, $\vec{\mu} \| \vec{\Omega}$. In
this case the GJ charge density in the polar cap of pulsar is negative
and there are electrons, which flow away from the polar cap. The
poloidal current $S$ in the open field line zone is negative (see
definition of the poloidal current eq.~(\ref{eq:B})). In the case of
anti-aligned rotator, i.e. $\vec{\mu}$ is antiparallel to
\vec{\Omega}, all signs of the physical quantities related to the
charge and current should be reversed.

Calculations have been performed for the following values of
\XL{}:0.15; 0.2; 0.3; 0.4; 0.5; 0.6; 0.7; 0.8; 0.9; 0.95; 0.99; 0.992.
An unique solution has been found for each of the above \XL's. Let us
consider in details physical properties of the obtained solutions.

\subsection{Poloidal current}
\label{sec:poloidal-current}

\begin{figure}
  \centerline{\includegraphics[clip,width=\columnwidth]{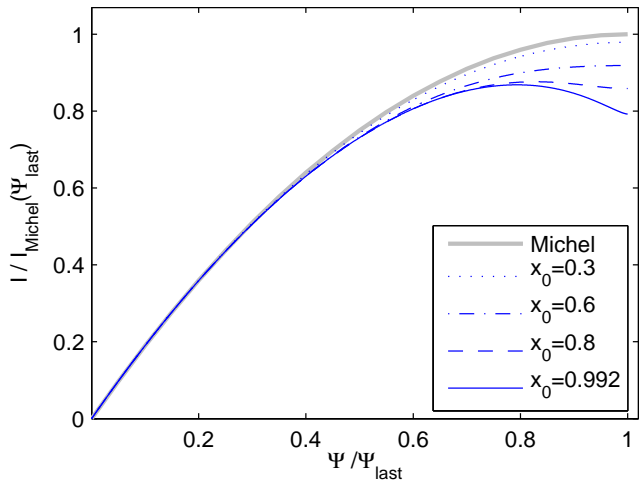}}
  \caption{Poloidal current distribution in the open field line zone
    normalised to the poloidal current from the corresponding Michel's
    solution. Inside the current sheet, for $\PsiL \le \psi \le
    \PsiL+d\psi$, (not shown here) the poloidal current decreases to 0}
  \label{fig:I}
\end{figure}

The poloidal current density $S$, calculated from the
formula~(\ref{eq:Spoloidal}), does not deviate for more than $\sim{}20$
per cents from the values given by the Michel's solution
\citep{Michel73}
\begin{equation}
  \label{eq:Spoloidal_Michel}
  S = - \psi \left(2-\frac{\psi}{\PsiL}\right)
  \,,
\end{equation}
see Fig~\ref{fig:I}. The smaller \XL, the smaller this deviation.  The
structure of the magnetosphere depends strongly on the poloidal
current distribution. In solutions with $\XL \goa 0.6$ there is a
domain in the open filed line zone, where \emph{volume} return current
flows. However, only a small part of the return current flows there,
the main part flows inside the current sheet. The size of this domain
gets smaller with decreasing of $\XL$, and for $\XL \loa 0.6$ the
return current flows only along the separatrix, see
Fig.~\ref{fig:Main}.  Qualitatively this property of solutions could be
explained as the following.  At the LC the
condition~(\ref{eq:CondAtLC}) is satisfied, so if $\pd_x\psi$ changes
the sigh the same occurs with the current term \SSp{}, and the
poloidal current density changes the sign.  Magnetic field lines close
to the null point are bend to the equatorial plane, but at large
distance they become radial.  So, for \XL{} close to 1 $\pd_x\psi<0$
for some field lines, and volume return current must flow along them.
When \XL{} decreases, more an more magnetic field lines at the LC will
be bend away from the equatorial plane until there will be no lines
bend to the equator. For field lines bend from the the equatorial
plane $\pd_x\psi>0$ and there is no volume return current along them.

A convenient representation of the current density in the closed field
line zone could be given by the current density distribution in the
polar cap \PC{j}. In our notations the current density in the polar
cap of pulsar normalised to the Goldreich-Julian current density
$\GJ{j}\equiv \Rgj c$ is given by (see appendix~\ref{sec:App-jpc},
eq.(\ref{eq:app-jpc-final}))
\begin{equation}
  \label{eq:jpc}
  \PC{j} = |\GJ{j}| \; \frac1{2}  
  S^\prime 
  ( 
    \left( \frac{\theta}{\PC{\theta}} \right)^2 \PsiL 
  )
  \,,
\end{equation}
where $\theta/\theta_\mathrm{pc}$ is the colatitude normalised to the
colatitude of the polar cap boundary $\theta_\mathrm{pc}$, it is
connected to the function $\psi$ through the relation
$\theta/\theta_\mathrm{pc}=\sqrt{\psi/\PsiL}$. In Fig.~\ref{fig:jpc}
\PC{j} is shown for several solutions with different \XL's. The current
density never exceeds the corresponding GJ current density and goes to
zero at the polar cap boundary. The latter property is the consequence
of the assumed magnetosphere's topology.  Indeed, from the condition
at the LC, eq.(\ref{eq:CondAtLC}), the current density along a given
magnetic surface is proportional to the partial derivative $\pd_x\psi$
at the LC, but in configurations with Y null point $\pd_x\psi =0$ for
$\psi=\PsiL$. The deviation of the current density $\PC{j}$ from the
GJ current density increases close to the polar cap boundaries
with increasing of \XL{}. For solutions with $\XL\goa0.6$
the current density \PC{j} changes the sign at some point near the
boundary. On the other hand, \PC{j} never exceeds the corresponding
Michel current density and approaches $j_\mathrm{Michel}$ when \XL{}
decreases.

\subsection{Drift velocity and force-free approximation}
\label{sec:drift-velosity}

The drift velocity in our notations is given by
\begin{equation}
  \label{eq:Ud(psi)}
  u_\mathrm{D} \equiv \frac{|\Ud|}{c} = 
  \frac{\Omega\varpi}{c} \frac{\Pol{B}}{B} =
  \frac{x}%
  {\sqrt{\displaystyle 1+\frac{S^2}{(\pd_x\psi)^2+(\pd_z\psi)^2}}}
  \,,
\end{equation}
\Pol{B} is the poloidal component of magnetic field.  The light
surface, i.e. the surface where the force-free approximation breaks
down coincide with the surface, where $u_\mathrm{D}=1$.  We verified
the applicability of the force-free approximations in each case. For
most of the cases calculations have been performed in the domain with
$x_{max}=8,z_{max}=7$, but for $\XL=0.2;0.7;0.992$ we performed
calculations also with $x_{max}=16,z_{max}=14$. In all cases the light
surface is located somewhere outside of these domains, see
Fig.~\ref{fig:Main}. The drift velocity distribution for solutions
with \XL{} close to 1 even at large distances from the null point
differs significantly from one in corresponding Michel's solution (the
solution with the same \PsiL), where the drift velocity is the
function of only $x$-coordinate. On the other hand, when \XL{}
decreases, $u_\mathrm{D}$ approaches the values from the corresponding
Michel's solution.

%

\subsection{Charge distribution in the magnetosphere}
\label{sec:charge-distr-magn}

Goldreich-Julian charge density in the magnetosphere in our notations
is given by
\begin{equation}
  \label{eq:RhoGJ}
  \Rgj =\rho_0\;
  \frac{\displaystyle \SSp-\frac{2}{x}\, \pd_x\psi}{1-x^2}\, ,
  \qquad
  \rho_0 \equiv \frac{\mu}{4\pi\RLC^4}
  \,.
\end{equation}
Close to the rotation axis the GJ charge density is negative and with
increasing of the colatitude it becomes positive. While for solutions
with $\XL \ga 0.6$ the domain of positively charged plasma extends to
infinity, for solutions with smaller \XL's it becomes finite (cf.
plots for $\XL=0.2$ with other plots in Fig.~\ref{fig:Main}). The
reason for this is the following. At large distance from the light
cylinder magnetic field lines becomes radial, so $\pd_x\psi$ is always
greater than 0.  Hence, there only the term \SSp{} is responsible for
changing of the charge density sign. However \SSp{} for $\XL\loa 0.6$
never changes the sign, see left plots in Fig.~\ref{fig:Main}.  For the
same reason the volume return current always flows trough the
positively charged domain. Close to the NS it passes trough the layer
where charge density changes the sign, see right plots in
Fig.~\ref{fig:Main}. At this layer the so-called outer-gap cascade
should develop \citep[see e.g.][]{Cheng/Ruderman76,Hirotani04}.

\begin{figure}
  \centerline{\includegraphics[clip,width=\columnwidth]{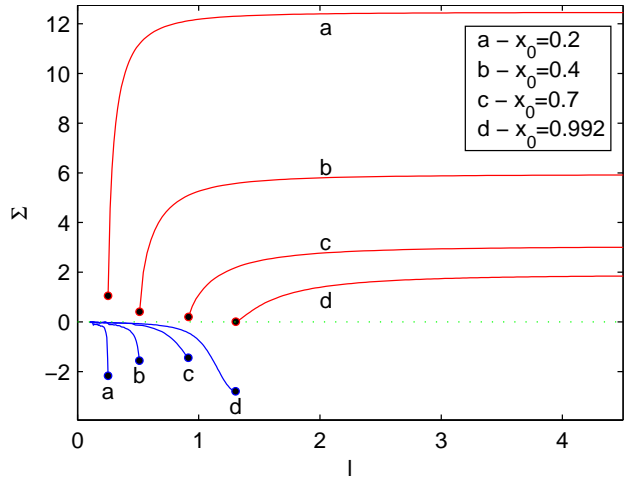}}
  \caption{$\Sigma$ -- linear charge density of the current sheet (see
    text) as a function of the distance $l$ \emph{along} the current
    sheet. $\Sigma$ is normalised to $0.5\,\mu/\RLC^2$. $l$ is
    measured in units of \RLC{}. The points marks the position of the
    corresponding null point. Note the jump in the charge density at
    these points. The dotted line corresponds to $\Sigma=0$.}
  \label{fig:sigma} 
\end{figure}

The force-free solution fixes not only the volume charge density, but
also the charge density of the current sheet.  As the electric field
at opposite sites of the current sheet is different, the current sheet
must have nonzero surface charge density.  In Fig.~\ref{fig:sigma} we
plotted the linear charge density $\Sigma$ of the current sheet as a
function of distance $l$ along the separatrix
\begin{equation}
  \label{eq:eta}
  \Sigma \equiv 2\pi \varpi \sigma
  \,,
\end{equation}
where $\sigma$ is the charge density of the current sheet.  $\Sigma$
represents the total charge of a volume co-moving with particles
flowing along the separatrix with the constant speed, emitted at the
same time (either at the NS or at ``infinity''). $\Sigma\equiv
\mathrm{const}$ would imply a constant velocity flow of particles of
\emph{one} sign. However for each solution $\Sigma$ is non-monotonic
function with discontinuity in the null point.  Such complicated
dependence of $\Sigma$ on $l$ implies some non-trivial physics
connected with particle creation in the current sheet, which is
discussed in the next section.

This complicated dependence of the current sheet charge density is
easy to understand if one consider the so-called ``matching
condition'' at the separatrix.  As it was shown by \citet{Lyubar90}%
\footnote{see also \citet{Okamoto74}, eq.~(69)}, %
at the current sheet the following condition for electric and magnetic
field in closed (c) and open (o) field line domains should be satisfied
\begin{equation}
  \label{eq:E2-B2}
  E^2_\mathrm{c} -  B^2_\mathrm{c} =  E^2_\mathrm{o} -  B^2_\mathrm{o}
  \,.
\end{equation}
This follows from integration of equation~(\ref{eq:PsrEq}) across the
current sheet. In the closed field line zone there is no toroidal
magnetic field. As it follows from eqs.~(\ref{eq:E}) and (\ref{eq:B})
the electric field 
\begin{equation}
  \label{eq:E=x_Bpol}
  E = x \Pol{B}
  \,.
\end{equation}
Substituting this equation into equation~(\ref{eq:E2-B2}) we get
\begin{equation}
  \label{eq:Bc_pol2-Bo_pol2}
  B^2_\mathrm{pol,\,c} - B^2_\mathrm{pol,\,o} =
  \frac{B^2_{\phi,\mathrm{o}}}{1-x^2}
  \,.
\end{equation}
From this and equation~(\ref{eq:E=x_Bpol}) follows that
$E_\mathrm{c}>E_\mathrm{o}$ and the charge density in the current
sheet between closed and open field line domain,
\begin{equation}
  \label{eq:sigma_cSH_OC}
  \sigma = \frac{1}{4\pi} (E_\mathrm{o}-E_\mathrm{c})
  \,,
\end{equation}
is always \emph{negative}. On the other hand, from the symmetry of the
system -- the electric field in regions $2$ and $2^\prime$ in
Fig.~\ref{fig:Topology-Y} has different directions, -- the charge
density of the current sheet in the open field line zone
\begin{equation}
  \label{eq:sigma_cSH_OO}
  \sigma = \frac{1}{2\pi} E_\mathrm{o}
\end{equation}
is always \emph{positive}

%
\begin{figure}
  \centerline{\includegraphics[clip,width=\columnwidth]{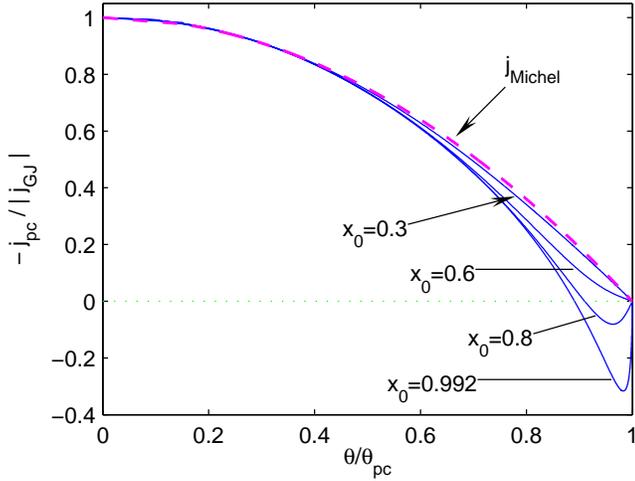}}
  \caption{Current density distribution in the polar cap of pulsar
    $\PC{j}$ as a function of the colatitude. $\PC{j}$ is normalised
    to the Goldreich-Julian current density $|\GJ{j}|$ and the
    colatitude is measured in units of the polar cap boundary
    colatitude \QPC.}
  \label{fig:jpc}
\end{figure}

%

%
\begin{figure}
  \centerline{\includegraphics[clip,width=\columnwidth]{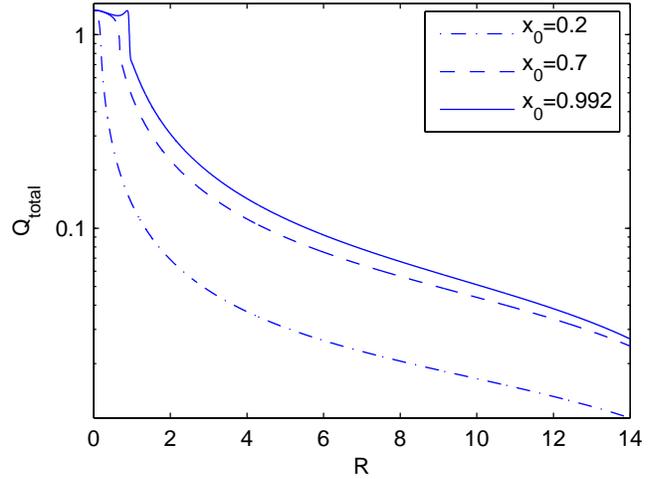}}
  \caption{The total charge inside the sphere of the radius $R$
    centred at the NS: charge of the NS + charge in the magnetosphere
    obtained by direct integration of \Rgj. $Q_\mathrm{total}$ is
    normalised to $0.5\,\mu/\RLC$. $R$ is measured in units of
    \RLC{}.}
  \label{fig:Qtotal}
\end{figure}

%
\begin{figure}
  \centerline{\includegraphics[clip,width=\columnwidth]{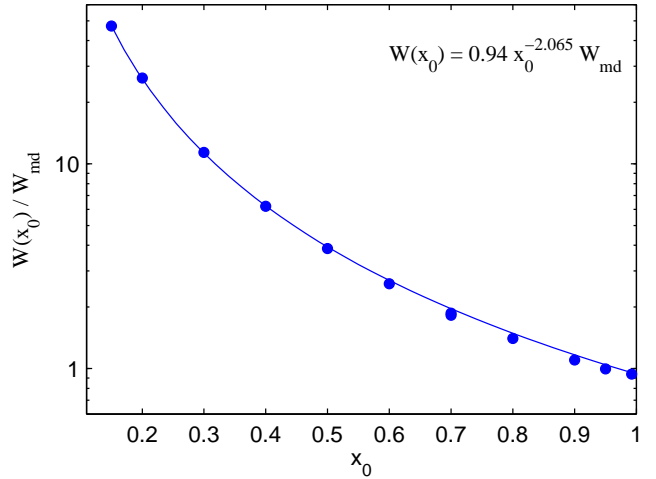}}
  \caption{Energy losses of the aligned rotator as a function of $x_0$
    in units of the corresponding magnetodipolar energy losses \Wmd.}
  \label{fig:W}
\end{figure}

%
\begin{figure}
  \centerline{\includegraphics[clip,width=\columnwidth]{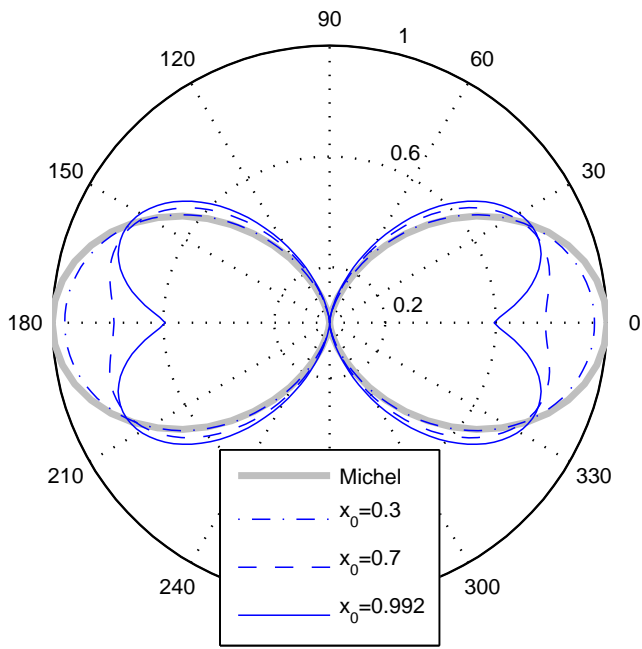}}
  \caption{Angular distribution of the energy flux $dW/d\omega$
    normalised to $-\PsiL^2\,\Wmd/(4\pi)$, see
    eqs.~(\ref{eq:dW/dOmega}) and (\ref{eq:dW/dOmega_Michel}).  The
    distributions shown here are taken at $R=4\RLC$ and correspond to
    their asymptotic forms, see text.}
  \label{fig:dwdo}
\end{figure}

The total charge of the system, i.e. the charge of the NS, the
magnetosphere and the current sheet together must be zero. The
boundary condition~(\ref{eq:bcond_outer}) implies that the total flux
of electric field through the sphere of a large radius is zero, hence
the total charge of the system must be zero. In Fig.~\ref{fig:Qtotal}
the total charge inside the sphere centred at the coordinate origin is
plotted as a function of its radius. The total charge of the system
goes rapidly to zero at large distances from the NS.  This plot could
be also considered as an additional test of the numerical procedure,
as the conservation of the total charge is not incorporated into the
numerical scheme.

\subsection{Energy losses}
\label{sec:energy-losses}

Energy losses of the aligned rotator in our notations are given by the
formula
\begin{equation}    
  \label{eq:W}
  W = \Wmd\; \int_{0}^{\PsiL} S\, d\psi
  \,,
\end{equation}
where $\Wmd$ is the absolute value of magnetodipolar energy losses,
here defined as
\begin{equation}
  \label{eq:Wmd}
  \Wmd = \frac{B_0^2 \RNS^6\Omega^4}{4 c^3}
  \,, 
\end{equation}
see appendix~\ref{sec:App-energy-losses}, eqs.~(\ref{eq:app-W}),
(\ref{eq:app-W_md}). In the obtained set of solutions $W$ is function
of \XL{}. With decreasing of \XL{} the amount of open magnetic field
lines increases and, as the poloidal current dependence on $\psi$ does
not changes substantially, the energy losses of aligned rotator
increases with decreasing of \XL, see Fig.~\ref{fig:W}.  Obtained
dependence of energy losses W on the position of the null point \XL{}
could be surprisingly well fitted by a single power law
\begin{equation}
  \label{eq:W(XLast)}
  W(\XL) \approx - 0.94\, \XL^{-2.065} \; \Wmd
  \,.
\end{equation}
This formula is similar to the one obtained from analytical
estimations using Michel current distribution (see
appendix~\ref{sec:App-energy-losses}, eq.(\ref{eq:app-W-theory}))
\begin{equation}
  \label{eq:W_theory(XLast)}
  W(\XL) \approx - \frac2{3} \XL^{-2} \; \Wmd
  \,.
\end{equation}

The angular distribution of the energy flux (see
appendix~\ref{sec:App-energy-losses}, eq.~(\ref{eq:app-dWdOmega}))
\begin{equation}
  \label{eq:dW/dOmega}
  \frac{dW}{d\omega} = \frac{\Wmd}{4\pi} \; 
  S \frac{\sqrt{x^2+z^2}}{x}
  \left( z\,\pd_x\psi - x\,\pd_z\psi \right)
  \,.
\end{equation}
In Fig.~\ref{fig:dwdo} this distribution is shown for several
solutions with different \XL{}. The Poynting flux distribution
quickly reaches its asymptotic form at distance from the null point of
the order of $1-2~\RLC{}$.  For example, in the case of $\XL=0.99$
distributions taken at $R=4$ and $R=14$ differs by no more than $\sim
3$ per cents.  For configurations with smaller \XL{}'s this deviation
is even less.  The smaller \XL{} the close the angular energy flux
distribution to the angular distribution in Michel's solution
\begin{equation}
  \label{eq:dW/dOmega_Michel}
  \frac{dW}{d\omega} =  - \frac{\Wmd}{4\pi}\;\PsiL^2 \sin^2\theta
  \,,
\end{equation}
because for small \XL{} the solution at large distances is very close
to the Michel's solution. In spite of recent works on modelling of
jet-torus structure seen in Crab and other plerions
\citep[see][]{Komissarov/Lyubarsky03,Bogovalov05}, we note that
magnetosphere configurations with larger \XL{} would stronger support
development of instabilities due to more asymmetric energy deployment
into the plerion, providing more pronounced disk structure.

%
\begin{figure}
  \centerline{\includegraphics[clip,width=\columnwidth]{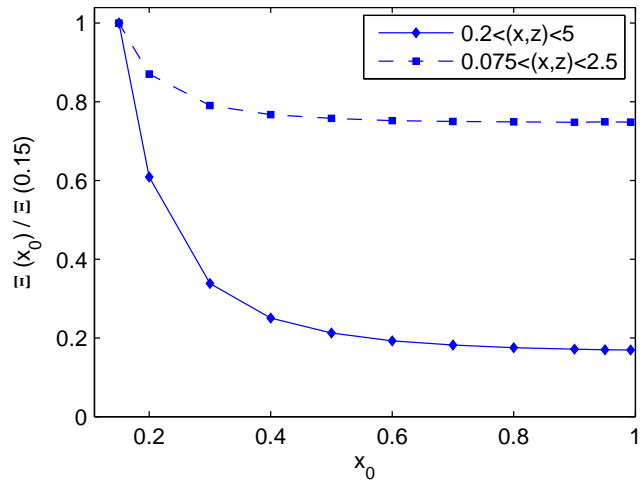}}
  \caption{Total energy of electromagnetic field in two different
    volumes of fixed sizes as a function of $x_0$. $\Xi$ is normalised
    to the corresponding value of $\Xi(\XL=0.15)$.}
  \label{fig:energy}
\end{figure}

\subsection{Total energy  of the magnetosphere}
\label{sec:total-energy-magn}

The total energy of electric and magnetic fields in the magnetosphere
$\Xi\equiv\int (B^2+E^2)/(8\pi)\: d\,V$ would give information which
configuration the system tries to achieve, the configuration with the
minimal possible energy. Obviously for the obtained solutions we could
calculate the energy only in a finite domain. Another problem is very
rapidly increase of magnetic field in the central parts, as
$r^{-3}$. As the magnetic field close to the NS is dipolar for each
configuration, we calculate the total energy in a domain excluding the
central parts. In order to verify the independence of the result on a
particular domain we calculate the total energy in the magnetosphere
in two different domains for each solution. These domains are defined
as
$0.2 \le x \le 5,0.2 \le z \le 5$ and
$0.075 \le x \le 2.5,0.075 \le z \le 2.5$ 
The results are plotted as a function of \XL{} in
Fig.~\ref{fig:energy}. The total energy of the magnetosphere increases
with decreasing of \XL{}, so the magnetosphere will try to achieve the
configuration with the maximal possible \XL{}.

%
\begin{figure}
  \centerline{\includegraphics[clip,width=\columnwidth]{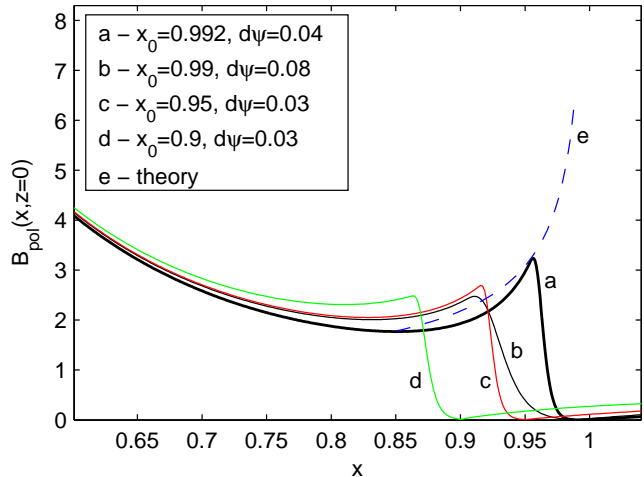}}
  \caption{Poloidal magnetic field strength in the equatorial plane as
    the function of $x$. $\Pol{B}$ is normalised to $\mu/\RLC^3$.  By
    the dashed line the theoretical prediction for $\Pol{B}(x,0)$ is
    shown for solution with $x_0=0.992$, eq.(\ref{eq:Binf}).}
  \label{fig:Binf}
\end{figure}

\subsection{Solution with $\XL \rightarrow 1$}
\label{sec:solution-with-xl}

The special case of $\XL \rightarrow 1$ has been considered by several
authors, because it was believed to be the real configuration of a
pulsar magnetosphere
\citep{Lyubar90,CKF,Uzdensky03,Gruzinov:PSR,Komissarov05}. This case
is peculiar in the sense, that magnetic field in the closed filed line
zone diverges in the Y null point. Indeed, from
equation~(\ref{eq:Bc_pol2-Bo_pol2}) it follows that near the null
point, when $\XL\rightarrow{}1$
\begin{equation}
  \label{eq:Binf}
  \Pol{B} \approx \frac{\mu}{\RLC^3} \; \frac{|S|}{\sqrt{2(1-x)}}
  \,.
\end{equation}
While the presence of the singularity was noted by \citet{Lyubar90}
and \citet{Uzdensky03}, \citet{Gruzinov:PSR} firstly realised that
such singularity is admitted, as it does not lead to the infinite
energy of magnetic field in the region surrounding the null point. In
Fig.~\ref{fig:Binf} the strength of the poloidal magnetic field along
the $x$-axis is plotted for different solutions. By the dashed line
the relation~(\ref{eq:Binf}) is shown. We see that when \XL{}
approaches the LC the magnetic field inside the closed zone begins to
grow close to the null point. This increase is more pronounced when
the thickness of the current sheet decreases. Agreement between the
curve for $\XL=0.992,d\psi=0.4$ and the dashed line is quite good.

\citet{Gruzinov:PSR} solved an equation for the separatrix in the
vicinity of the null point $\XL=1$ and have found that the angle at
which separatrix intersects the equatorial plane should be
$77.3\degr$. In our calculations we found this angle to be $\approx
78\degr$ for $\XL=0.992, d\psi=0.04$ and $\approx 70\degr$ for
$\XL=0.99,d\psi=0.08$.  So, our numerical solution shows good
agreement with the analytical one.  Energy losses found by
\citet{Gruzinov:PSR} are $1.0\pm 0.1$, what quite good agrees with
values for $W$ from Table~\ref{tab:SolProp}. Value of $\PsiL=1.23$
calculated by \cite{Contopoulos05} coincide with ones from
Table~\ref{tab:SolProp} and is close to $\PsiL=1.27$ obtained by
\citet{Gruzinov:PSR}, although both of these results have been obtained
with codes having worse numerical resolution than the code used in
this work.

\section{Discussion}
\label{sec:discussion}

It seems naturally to assume, that force-free configurations are
energetically preferably in comparison with configurations where there
are geometrically large volumes with parallel electric field%
\footnote{however see e.g.  \citet{Smith/Michel01,Petri/Heyvaerts/02}}.
Accepting this, we conclude that magnetosphere of a pulsar should
evolve through a set of force-free configurations. It does not
necessary mean that for a relatively short transition time the system
could not be essentially non-force-free, but rather that the most time
the magnetosphere of an active pulsar is force-free.

\subsection{Polar cap cascades and force-free  magnetosphere}
\label{sec:role-polar-cap}

In a force free configuration the current density distribution is not
a free parameter, it is set by the structure of the magnetosphere, for
example, by the value of \XL{} in the case of Y-configuration. However,
the current in the magnetosphere of pulsar is supported by
electron-positron cascades in the polar cap, i.e. the most of current
carriers are produced in the magnetosphere and are not supplied from
external sources.  Independently of neutron star crust properties,
i.e. whether or not charged particles could be extracted from the
surface, in polar cap of young pulsars electron-positron cascades are
developed filling the magnetosphere of the star with particles
\citep{Ruderman/Sutherland75,Arons/Scharlemann/78,Muslimov/Tsygan92}.
Also these particles are necessary in order to support MHD like
structure of the magnetosphere. The current in the magnetosphere flows
trough this cascade region, hence, the cascade, which properties
depend on local magnetic field structure, has to adjust to the global
properties of the magnetosphere too, namely to the current density
flowing through it. We focus here on the case of stationary cascades.
The hypothesis about stationarity of the polar cap cascades, when
temporal variations of the accelerating electric field over the whole
polar cap is much less than the accelerating field itself is widely
adopted \citep[e.g.][]{Daugherty/Harding82,Ruderman/Sutherland75}.  We
briefly address also the case of essentially non-stationary cascade
\citep{Levinson05}.

As it was shown by \citet{Lyubar92}, for current adjustment in the
stationary cascades a particle inflow from the magnetosphere into the
cascade region is required. The typical current density,
self-consistently supported by stationary polar cap cascades, is close
to $\GJ{j}$. For current densities, both larger or smaller, than the
Goldreich-Julian one, a particle inflow is necessary.  The source of
inflowing particles needed for current adjustment could be outer gap
cascades, operating at the surface where GJ charge density changes the
sign \citep{Cheng/Ruderman76}.  On the other hand, inflowing particles
could be provided by the pulsar wind, where some outflowing particles
could be reversed back to the NS due to momentum redistribution or due
to small residual electric field arisen as the magnetosphere tries to
support a force-free configuration.  However, the zone where particles
could flow \emph{toward} the NS is limited by the light cylinder (see
Appendix~\ref{sec:Ux_max}).  So, the source of inflowing particles
must be inside the LC.

For \citet{Ruderman/Sutherland75} cascades, when particles can not be
extracted from the NS surface and are produced in the discharge zone,
the adjustment mechanism works as follows. Inflow of positrons
increases the current density, inflow of electrons decreases it.  In
the first case the inflowing positrons decrease charge density in the
Pair Formation Front (PPF) and more electrons is necessary to adjust
the charge density to the GJ value.  This additional electrons
together with inflowing positrons increase the current density. When
there is an inflow of electrons, less primary electrons are necessary
in order to support the GJ charge density at the PFF. Inflowing
electrons are turned back at the PFF, and compensate the inflowing
electric current.  The outflowing current is only due to the primary
electrons from the discharge zone, so the current density is less than
$\GJ{j}$.

If particles could almost freely escape from the NS crust, the pulsar
operates in the so-called Space Charge Limited Flow (SCLF) regime and
the current density can not be essentially less then $\GJ{j}$. Indeed,
the charge density in the discharge region, below PFF, is close to
$\Rgj$ and accelerating electric field forces charges to
outflow with relativistic velocities
\citep{Arons/Scharlemann/78,Muslimov/Tsygan92}. For cascades operating
in SCLF regime the mechanism of current adjustment works similarly for
inflowing positrons.  The particle inflow could \emph{increase} the
current density, but not decrease it. Only when the accelerating
electric field is almost completely screened, the current density could
be significantly less than $\GJ{j}$. However, in order to screen this
accelerating field, charged particles inflowing from the magnetosphere
must penetrate practically up to the NS surface, i.e. they must have
Lorentz factors comparable to the Lorentz factors of particles
accelerated in the polar gap. In other words, somewhere in the
magnetosphere inside the LC there should be zone(s) where particles are
accelerated as effective as they would be accelerated in in the polar
cap. Either the accelerating field there should be comparable to the
one in the polar cap or the size of this zone would be essentially
larger than some NS's radii.  Both seems to be inappropriate.

In both of these cases in order to support volume \emph{return}
current, flowing in the direction opposite to $\GJ{j}$, the
accelerated field in the polar cap discharge zone must be completely
screened and the particles filling the magnetosphere along magnetic
field lines with return volume current must be produced somewhere in
the magnetosphere.  The accelerating electric field in the polar cap
zone, being proportional to the magnetic field strength, is much
stronger than any possible accelerating electric field far from the
NS. Hence, the presence of the return volume current in the force-free
magnetosphere seems to be incompatible with the force-free
configurations of the magnetosphere, because the acceleration of
particles to the required Lorentz factors with much weaker electric
field requires large non-force-free domain(s) in the magnetosphere.
The situation with non-stationary cascades is poor investigated,
currently there is only one work dedicated to detailed studies of
significantly non-stationary cascades -- \citet{Levinson05}. However,
we see no way how it would be impossible to support both particle
production in the polar cap cascade and an average current having
opposite direction to the direction of accelerating electric field,
see also \citet{Arons79}.

In our consideration we assumed that the GJ charge density in the
polar cap does not deviate substantially from its canonical value
\citep{GJ}
\begin{equation}
  \label{eq:RhoGJ_PC}
  \Rgj = -\frac{\Omega B_0}{2\pi c}
  \,.
\end{equation}
This is the case when the boundary of the polar can be considered as
equipotential, i.e. having very high conductivity. However if its
conductivity is very low and surface charge density distribution at
separatrix in the polar cap is different from the one in force-free
solution, the GJ charge density can substantially deviate from values
given by formula~(\ref{eq:RhoGJ_PC}). In this case the characteristic
current density flowing trough the cascade region would be different
from the canonical value of $-(\Omega B_0)/(2\pi)$ and, in
principal, it could approach the values required by the global
magnetospheric structure, i.e. the problem of current adjustment could
be solved by modifying \Rgj{} instead of adjusting the deviation of
$j$ from $\GJ{j}$. Let us analyse this possibility. The largest part
or the whole return current flows along the separatrix. It could be
electrons returning from region behind the light surface%
\footnote{current sheet is not a force-free domain and considerations
  from Appendix~\ref{sec:Ux_max} are not valid here} %
of ions outflowing from the NS surface \citep[see
e.g.][]{Spitkovsky/Arons04}.  If there are electrons in the current
sheet close to the NS, then substantial deviation of electric field
from the force-free value will give rise to electron-positron cascades
producing enough particles to make separatrix near equipotential. Only
ions, which much hardly emit photons capable to produce
electron-positron pairs could support essentially non-equipotential
polar cap boundary. However, as it was mentioned before, each
particularly force-free configuration fixes the surface charge density
distribution along the current sheet everywhere where it is
applicable. Independently on detailed structure of the polar cap zone
the surface charge density along the separatrix between closed and
open field lines is \emph{negative}, i.e. it must be enough electrons
there, or the magnetosphere will be not force-free, see
section~\ref{sec:charge-distr-magn} and Fig.~\ref{fig:sigma}.
Although in the discharge zone above the polar cap force-free
approximation is not valid, and arguments of
section~\ref{sec:charge-distr-magn} cannot be directly applied to the
current sheet at the polar cap boundaries, electrons must be there for
the following reason. The current sheet is a region where force-free
approximation is broken, at least in some places, for example in the
null point, where the surface charge density is discontinuous. As the
return current flows in the current sheet, the parallel electric field
will be directed from the NS, accelerating electrons in the current
sheet toward the NS surface.  Hence, in the current sheet at the polar
cap boundary there are electrons too and this boundary will be
approximately equipotential.  Consequently, the GJ charge density in
the polar cap should be close to the canonical
value~(\ref{eq:RhoGJ_PC}) and in order to support a force-free
configuration of the magnetosphere a current adjustment mechanism is
necessary.

For current adjustment high particle density in the magnetosphere is
required. Indeed, only a small fraction of all particles could be
turned back to the NS. There must be enough inflowing particles for
adjusting of the current density in the polar cap, i.e. its number
density should be of the order of $\Rgj/e$.  Hence, the particle
number density in the magnetosphere must be $\gg\Rgj/e$. However
almost all particles in the magnetosphere are produced in the polar
cap and outer gap cascades, and a rather complicated coupling between
cascade regions and pulsar magnetosphere arises. The weaker the
cascades, the less particles are produced there, the smaller deviation
from the GJ current density could be supported.  Hence, when pulsar
becomes older, the number of particles created in polar cap and outer
gap cascades is smaller and the maximal deviation of the current
density from $\GJ{j}$ will be smaller. If the magnetosphere remains
force-free, its configuration must be changed in order to adjust to
the new allowed current density. However, this new configuration would
result in different energy losses of the pulsar, i.e. the ratio of the
real losses to the losses given by the magnetodipolar formula will be
different from the same ratio in previous configuration. So, generally
speaking, the evolution of pulsar angular velocity derivative will not
follow the power law $\dot{\Omega}\propto-\Omega^3$, as it is
predicted by the magnetodipolar formula.

In the case of non-stationary cascades there are evidence that no
particle inflow into the cascade region may be necessary in order to
support current densities both larger and smaller than \GJ{j}
\citep[see][]{Levinson05}. However, for creation of ``wave-like''
pattern of accelerating electric field \citep{Levinson05}, necessary
for support of small current densities together with reasonable pair
creation rate, high pair density is required. With ageing of the
pulsar the maximal achieved electric field and pair density will
decrease and shorten the range of allowed current densities. This
would lead to the evolution of the magnetosphere similar to the case
with stationary cascades.

Arguments presented here are based on qualitative analysis of the polar
cap cascade properties. In order to make quantitative predictions a
more detailed investigation of polar cap cascades is
necessary regarding stationarity, ranges of current densities
supported without particle inflow from the magnetosphere, and
stability of the cascades in presence of particle inflow from the
magnetosphere.

\subsection{Configurations with Y null point}
\label{sec:conf-with-y}

%
\begin{figure}
    \includegraphics[clip,width=\columnwidth]{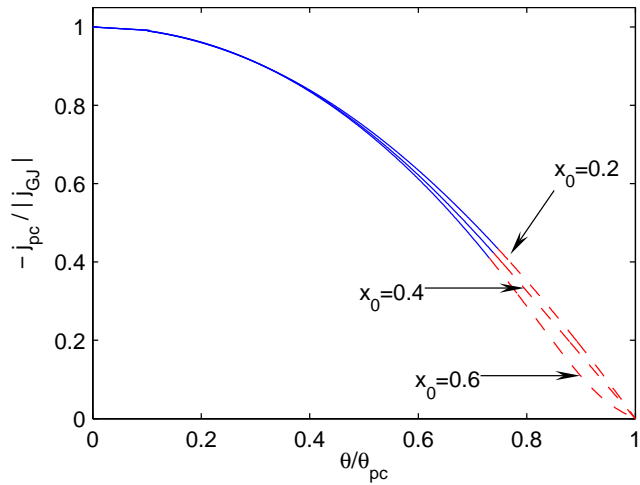}
    \caption{Current density distribution in the polar cap of pulsar
      $\PC{j}$ as a function of the colatitude. Normalisation of
      physical quantities is the same as in Fig.~\ref{fig:jpc}. The
      colatitude ranges where the current density deviation from
      \GJ{j} could be supported by particles produced in outer gap
      cascades are indicated by dashed lines. The current density at
      colatitudes where \PC{j} is shown by solid line should be
      supported by particles \emph{reversed} inside the light
      cylinder.}
  \label{fig:jpc_ogc}
\end{figure}

Let us analyse the behaviour of the magnetosphere of aligned pulsar
under assumption that the null point is always of Y type. Here again
we mean this in a time average sense, i.e. we neglect possible
non-stationary processes \citep[see
e.g.][]{Komissarov05,Contopoulos05} in the current sheet operating on
small scales ($\ll \RLC$), like building of small plasmoids. If
non-stationary variations of the current sheet remains small, the
stationary solution should adequately describe the properties of
magnetosphere.  The total energy of the magnetosphere decreases with
increasing of \XL, see Fig.~\ref{fig:energy}.  Apparently the system
will try to achieve the configuration with the minimum possible
energy, when $\XL=1$. However, when the restrictions set by the polar
cap cascades are taken into account the picture becomes more
complicated.

In solutions with Y null point the current density in the
magnetosphere close to the polar cap boundaries is always less than
the GJ current density, it does not exceed the Michel current density,
see section~\ref{sec:poloidal-current}, Fig.~\ref{fig:jpc}. The
current adjustment mechanism could adjust the current density to the
values \emph{less} than $\GJ{j}$ for stationary cascade model with no
particle escape from the NS surface, in \citet{Ruderman/Sutherland75}
model.  If pulsar operates in SCLF regime the current density can not
be essentially less then $\GJ{j}$.  Hence, force-free solutions with Y
null point are possible if the stationary polar cap cascade operates
in Ruderman-Sutherland regime, or if the cascade is significantly
non-stationary, the latter case however demands more detailed
investigations.  On the other hand, for solutions with $\XL\ga 0.6$,
the current density $\PC{j}$ close to the polar cap boundary has
different sign than $\GJ{j}$ and such force-free configurations are
probably never realised.

As it was mentioned in section~\ref{sec:role-polar-cap} the inflowing
particles could be produced either in the pulsar wind or in the outer
gap cascades.  The outer gap cascade could operate at the surface
where GJ charge density changes the sign \citep{Cheng/Ruderman76}.
Only relatively small amount of open field lines cross this surface.
Hence, particles produced in the outer gap cascades could not adjust
the current density along all magnetic field lines. On
Fig.~\ref{fig:jpc_ogc} we plot current density in the polar cap of
pulsar for several \XL{} and indicate by the dashed line the
colatitudes where particle inflow from the outer gap cascade would be
possible. The critical colatitude, where particle inflow from the
outer gap cascade is still possible, corresponds to the field line
with the smallest $\psi$ passing the surface of $\Rgj=0$ inside the
LC.  At other colatitudes reversed particles from the pulsar wind
(from inside the LC!)  are necessary in order to adjust the current
density.

In Fig.~\ref{fig:jpc} one can see that the deviation of the current
density from \GJ{j} although remaining large, becomes smaller with
decreasing of \XL. So, if the magnetosphere remain force-free, with
ageing of the pulsar, the configuration should change to the one with
smaller current density deviation from \GJ{j}. Hence, if the
force-free magnetosphere preserve its topology, with slow-down of the
neutron star the size of the closed field line zone becomes smaller.
Immediately consequence of this is the increasing of electromagnetic
energy losses respectively to the corresponding ``magnetodipolar''
energy losses according to equation~(\ref{eq:W(XLast)}), see
Fig.~\ref{fig:W}.  If at some time we approximate the dependence of
\XL{} on the angular velocity of NS rotation by the power law
\begin{equation}
  \label{eq:XL(Omega)}
  \XL \propto \Omega^\xi
  \,,
\end{equation}
where $\xi$ is in reality a (complicate) function of pulsar age.
$\xi>0$ because \XL{} decreases when pulsar became older. Substituting
it into the formula for pulsar energy losses (\ref{eq:W(XLast)}) we
get
\begin{equation}
  \label{eq:W(Omega)}
  W \propto \Omega^\alpha,\qquad \alpha=4-2.065\,\xi
  \,,
\end{equation}
and for pulsar braking index
\begin{equation}
  \label{eq:n_break}
  n = \frac{\ddot{\Omega}\Omega}{\dot{\Omega}^2} = \alpha-1 =
  3-2.065\,\xi
  \,,
\end{equation}
i.e. the breaking index is always less than 3!

Let us speculate that configurations with Y null point are
energetically preferable over all possible solutions (force-free and
non-force-free ones) and the polar cap cascade operates in
Ruderman-Sutherland regime. Then as long as particles produced in the
cascade regions will be able to support necessary current densities
the pulsar magnetosphere should evolve with time as described above,
and at each moment of time the configuration should be stable. Indeed,
due to reconnection of open field lines in the equatorial current
sheet the magnetosphere tries to achieve the energetically most
preferably configuration, with $\XL=1$, but weaker cascades could not
inject enough particles into the magnetosphere and support larger
deviation of the current density from $\GJ{j}$. So, \XL{} at each
moment of time correspond to the configuration with current
distribution having the maximal possible deviation from $\GJ{j}$. The
polar cap cascade zone is \emph{the part} of the whole system which do
not allow closed field zone to have the maximal possible size.
Conclusions of \citet{Spitkovsky05_poland}, \citet{Komissarov05} and
\citet{McKinney:NS:06} about instability of all configuration with
$\XL<1$ is the result of assumption about possibility of arbitrary
current density distribution in the pulsar magnetosphere.

We should note also another peculiarity in Y-configuration --
the jump in the surface charge density along the separatrix in the
null point, where the charge density changes the sign, see
section~\ref{sec:charge-distr-magn} and Fig.~\ref{fig:sigma}. The
return current flows along the separatrix: electrons to the NS,
ions/positrons from the NS. The surface charge density in the current
sheet has different sign before and after the null point.  We note
that there is a jump in the charge density, not just a continuous
changing of the charge density like it takes place across the surface
where $\Rgj=0$.  What happens in the null point, how such jump in the
charge density could be supported when there is a continuous particle
flow carrying the return current? Do an electron-positron cascade
operates here?  Both the magnetic field and soft X-ray radiation of
the NS are too weak here and electron-positron pair creation is
suppressed.  Anyway the problem requires additional investigations.

%
\begin{figure*}
    \includegraphics[clip,width=\columnwidth]{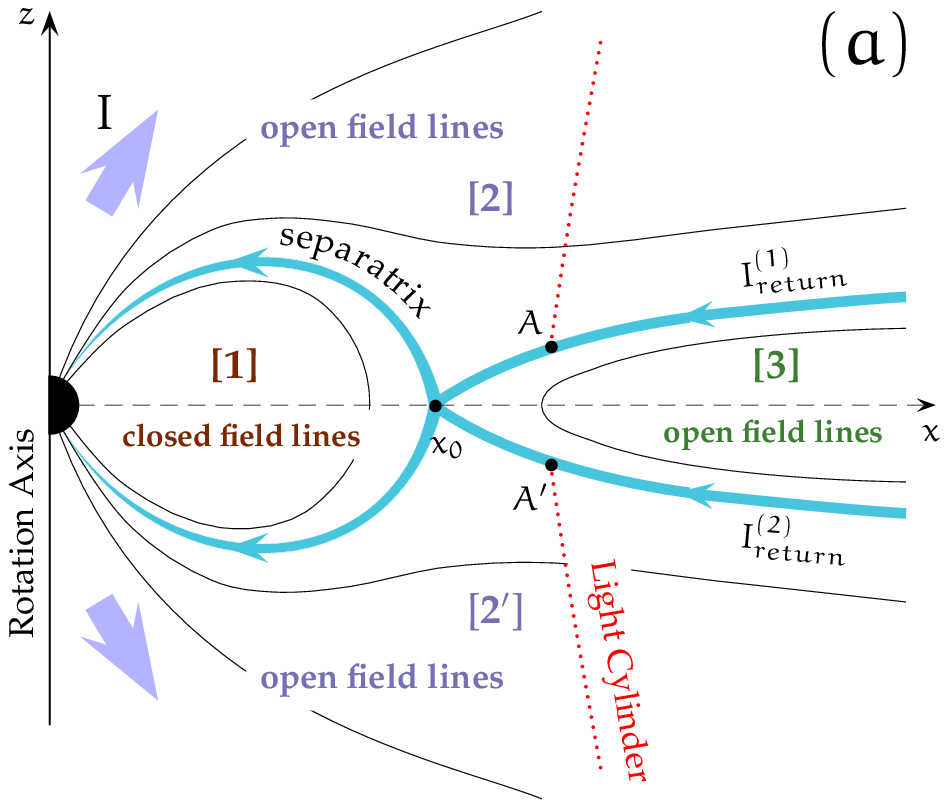}
    \qquad
    \includegraphics[clip,width=\columnwidth]{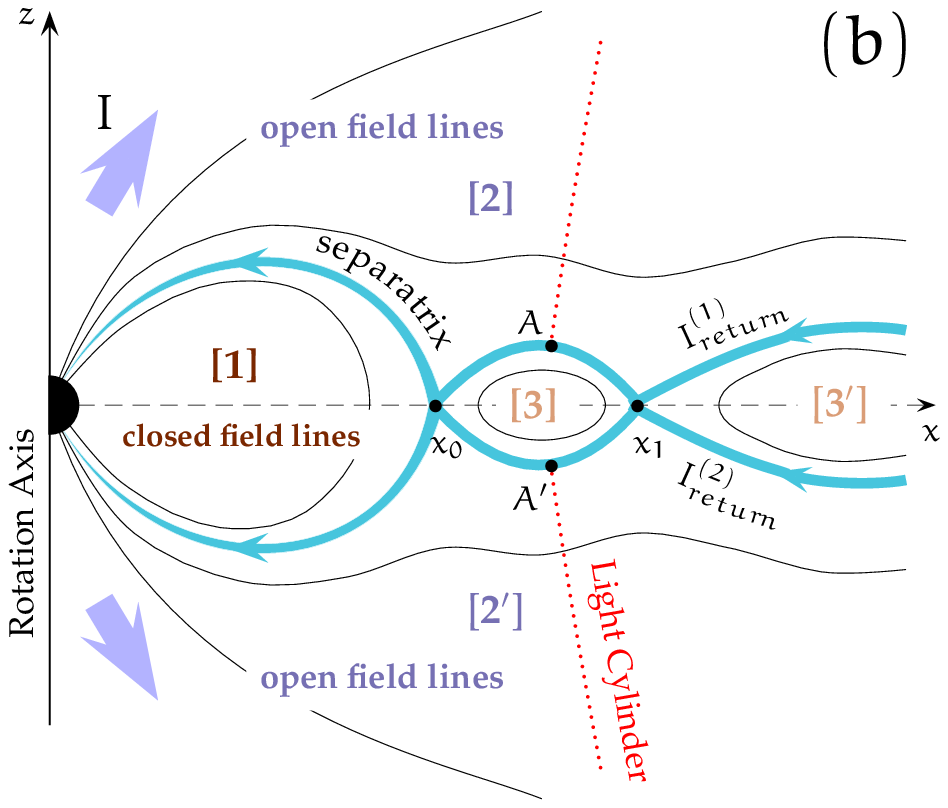}
    \caption{Configurations of magnetic field in the magnetosphere of
      aligned rotator with X-null point.%
      \textbf{(a)} -- after the null point \XL{} the separatrix goes
      away from the equatorial plane and never intersects it. There
      are 3 regions with open magnetic field lines: $[2]$,$[2^\prime]$
      and $[3]$.%
      \textbf{(b)} -- after the null point \XL{} the separatrix goes
      away from the equatorial plane but than intersects it (several)
      times in points $x_1,\dots$. There
      are several regions with closed magnetic field lines: $[1]$,$[3]$
      and $[3^\prime]$.}
  \label{fig:Topology-X}
\end{figure*}

\subsection{Alternatives to force-free Y configurations}
\label{sec:y-vs.-x}

In the force-free magnetosphere with Y-like null point the deviation
of the current density from \GJ{j} is always
large~$\la|\GJ{j}-j_\mathrm{Michel}|$, especially close to the polar
cap boundaries. For older pulsar with weak cascades it would be
problematic to adjust the current flowing through the polar cap to the
required value.  On the other hand, even if force-free
Y-configurations are energetically more preferably over all possible
solution, the stationary polar cap cascade operating in SCLF regime
does not allow current density necessary to support an Y-configuration
even in young pulsars.  In that cases the magnetosphere could become
non-force-free.

The possible alternative to a magnetosphere, becoming non-force-free
already at the distance of the order of \RLC{} from the NS, would be
a force-free magnetosphere with X-like null point.  The current
density deviation from the GJ current density in the magnetosphere
with X-like null point would be less that in the Y-configuration for
the following reason. In X-configuration with $\XL{}<1$
condition~(\ref{eq:CondAtLC}) must be satisfied in points at the light
cylinder above $A$ and below $A^\prime$, see
Fig.~\ref{fig:Topology-X}.  In that points $\pd_x\psi>0$, consequently
$\SSp(\psi\le\PsiL)>0$ and $j(\psi\le\PsiL)<0$ everywhere, neither
changing sign nor approaching zero. In principle, for \XL{} not to
close to the light cylinder and points $A$, $A^\prime$ not too close
to the equatorial plane the deviation of the current density from
$\GJ{j}$ could be made rather small allowing even week cascades to
support the current density, because in this case only small
correction to the current density would be necessary.  It is not clear
now do force-free X-configurations exists and how they would look like
(this work is in progress), however there is no clear physical reason
forbidding such possibility.

For X-configuration the jump in surface charge density of the current
sheet in the null point could be  avoided or at least reduced in
magnitude. Considerations from
section~\ref{sec:charge-distr-magn} can be applied to the current
sheet separating regions $2(2^\prime)$ and 3 in
Fig.~\ref{fig:Topology-X}. If directions of the poloidal magnetic
field in these regions coincides%
\footnote{The current sheet in such configuration is necessary,
  because in the force-free magnetosphere charged particles can not
  flow across magnetic field lines, so after the null point they
  should flow along a very thin layer too.}, %
the charge density in this current sheet could be negative too.
Indeed, the electric field in regions 2 ($E_2$) and 3 ($E_3$) close to
the separatrix in that case has the same direction. If $E_3>E_2$, the
charge density of the current sheen is negative and for such
configuration the charge density at separatrix does not change the
sign, but even if $E_2>E_3$ the positive charge density would be less
then in the case of equatorial current sheet, when the electric field
has to change the sign.  If the directions of poloidal magnetic field
in these regions are different, the same problem with the current
sheet as in Y-configuration arises for any values of the electric
field inside the domain 3.

X-configuration was criticised \citep{Lyubar90} because in the
force-free case there is no source of external magnetic field, which
could fill the region 3 in Fig.~\ref{fig:Topology-X}. However, ones
formed, this region may be supported by the global current system in
the magnetosphere.  Reconnection in the equatorial current sheet
\citep{Komissarov05,Contopoulos05} could lead to instantaneous
formation of magnetic loops, which would grow in size and form some
kind of X-configuration. These loops would try to merge with the
closed field line zone or fly away, but current distribution in the
force-free magnetosphere will not support such configurations, so the
resulting X-configuration could be stable. How would force-free
solutions with X null point look like is not clear now, may be there
could be solutions with many X null points, i.e. when there are
several islands of closed field lines along equatorial plane (see
Fig.~\ref{fig:Topology-X}(b)), but such complicated system may be
unstable.  On the other hand, the configuration shown in
Fig.~\ref{fig:Topology-X}(a)%
\footnote{such configuration has been considered in several works
  \citep[e.g.][]{GurevichBeskinIstomin_Book,Beskin/Malyshkin98}}
can not be entirely force-free. Indeed, there is no poloidal current
in zone~[3], as it has to have the same direction in both hemispheres,
hence, the magnetic field there is purely poloidal.  This implies that
the plasma here does not rotate and there is no currents in the
force-free case which could generate the magnetic field. However, we
could speculate that in zone~[3] at some (large) distance from the
null point, where force-free approximation is broken, a system of
currents is build up, which generate the magnetic field also in the
force-free domain of zone~[3], close to the null point. In this case
the magnetosphere may be force-free at several sizes of the LC.

Time-dependent simulations of aligned rotator magnetosphere probably
could clear what kind of configuration is realised.  Simulations of
\citet{Spitkovsky05_poland,Komissarov05} and \citet{McKinney:NS:06} do
not incorporate restriction on the electric current in the
magnetosphere due to polar cap cascade. In time-dependent codes the
restriction set by the polar cap cascades should be formulated in form
of boundary conditions on the poloidal current density. This could be
done, for example, by introduction of an artificial Ohm's law along
open field lines in the ``polar cap'' of pulsar, i.e. in regions close
to the NS surface, like
\begin{equation}
  \label{eq:Ohm_cascade}
  \vec{j}=\max( \sigma_{||}\vec{E}, \,\vec{j}_\mathrm{cas} )
  \,,
\end{equation}
where $\vec{j}_\mathrm{cas}$ corresponds to the minimal possible current
density along the particular field line allowed by cascades.
$\sigma_{||}$ is conductivity along magnetic field lines, specific to
each particular code. In order to set these restriction the knowledge
of polar cap cascade properties is necessary, what is another very
complicated problem.

The possibility of non-stationary magnetosphere also can not be
excluded at the current state of research. In order to decide between
possible configurations (e.g. force-free with X or Y null points,
non-force-free stationary magnetosphere or significantly
non-stationary configurations) more detailed studies of polar cap
cascades and stability of the current sheet are necessary.

\section{Conclusions}
\label{sec:conclusions}

We have studied in details stationary configurations of the force-free
magnetosphere of aligned rotator with Y null point.  Assumption about
Y-configuration of the magnetosphere is very popular and this case had
demanded careful investigations.  We find a set of force-free
solutions parametrised by the position of the neutral point \XL{}.
Results presented in this work for $\XL=1$ agree very well with ones
obtained by other authors
\citep{Contopoulos05,Gruzinov:PSR,Komissarov05,Spitkovsky05_poland}.
We calculated physical characteristic of obtained solutions, and
analysed properties of force-free magnetosphere with Y-like null
point. For solutions with \XL{} close to 1, we found that despite
similarly distributed magnetic surfaces at large distances from the
light cylinder, they differs substantially from the split monopole
solution of \citet{Michel73} regarding distribution of physical
quantities (drift velocity, energy flux distribution, etc.). When the
null point lies well inside the light cylinder, solutions approach the
Michel one, the agreement being better for smaller values of \XL's.

We analysed the role, which cascades in the polar cap play in
formation of the overall structure of the magnetosphere. Although its
properties depends mostly on the local physics in the polar cap of
pulsar, this cascade region sets serious limitation on the current
density in the whole magnetosphere. In some sense the non-trivial
physics of the cascade plays a role of complicated boundary conditions
for MHD equations describing structure of the magnetosphere --
arbitrary current density are not allowed. Changes in boundary
conditions influence the whole solution. We argue than not all
possible Y-configurations can be realised. Moreover, the restrictions
set by some cascade models questions the existence of stationary
force-free Y-configuration.  To our opinion, there are two problems
with the force-free magnetosphere in Y-configuration: (i) current
density strongly deviates from the GJ current density (ii) the charge
density along the current sheet has discontinuity and changes the
sign. These problems could be avoided in stationary force-free
X-configurations.

We argue that with ageing of pulsar and decreasing of the cascade
power, the magnetosphere must involve with time, i.e.  it should
change to the configurations where the deviation of the current
density from $\GJ{j}$ will be smaller. In the case of force-free
Y-configuration the closed field line zone grows slower that the light
cylinder during pulsar slowdown.  This leads to \emph{decreasing} of
pulsar breaking index below the value 3, predicted by the
magnetodipolar formula. This effect is present in aligned rotator
because of the current adjustment in the polar cap of pulsar. Similar
behaviour should be present in configurations with X-like null point
too. In analytical model of \citet{Beskin/Malyshkin98} it was shown
that the minimum energy of the magnetosphere is achieved when \XL{}
approaches the light cylinder. Although their model is an
oversimplification of the real problem, this results should be
qualitatively true.  X-configuration with $\XL=1$ requires strong
deviations of $j$ from $\GJ{j}$, so with ageing of pulsar \XL{} should
decrease, increasing number of the open field lines, leading to
increasing of energy losses and decreasing of pulsar breaking index.
This should be true also for inclined rotator, at least for not very
large inclination angle. Recently \citet{Spitkovsky/Contopoulos05}
proposed another explanation for breaking index of pulsar being less
than 3. In their model it is caused also by shrinking of the closed
field line zone, but they assumed this is due to differences in
characteristic time, at which magnetosphere reaches the new
configuration due to reconnection of new magnetic field lines, and the
time of increasing of the light cylinder radius. However the reason of
such slow reconnection in the current sheet is not clear.

Obtained solutions could be used for comparison with observations. For
example, magnetic field lines are differently twisted in different
solutions, this could be compared with magnetic field geometry
inferred from pulsar polarisation measurements \citep[see][and
references there]{Dyks/Harding04}. However, the real pulsar
magnetosphere could be non-stationary and/or non-force-free, and this
issue could be verified only within more detailed time-dependent
approach. On the other hand, in the latter case observational
manifestations of pulsar will be significantly different from models
with force-free magnetosphere, where main emission comes from polar
cap and from the outer gap zones.

\section*{Acknowledgments}

I thank V.~Beskin for numerous fruitful discussions, helpful
suggestions and encouraging comments.  I'm grateful to A.~Spitkovsky
for fruitful discussions and useful comments on the draft version of
this article.  I acknowledge K.~Hirotani, Yu.~Lyubarskij, and
J.~P\'etri for discussions.
I wish to thank anonymous referee for helpful critical comments.
I thank Max-Planck-Institut f\"ur Kernphysik (Heidelberg, Germany) for
hospitality.  This work was partially supported by German Academic
Exchange Service (DAAD) grant A/05/05462, and RFBR grant 04-02-16720.

\bibliographystyle{mn2e} \bibliography{psreq}

\appendix
\section[]{Current density in the polar cap}
\label{sec:App-jpc}

The poloidal current density is given by \citep[see][]{BeskinBook}
\begin{equation}
  \label{eq:app-jpc-general}
  \Pol{\vec{j}} = \frac{\D I \times \Uvec{\phi}}{\varpi} =
  \frac{d I}{d\Psi} \Pol{\vec{B}}
  \,,
\end{equation}
the latter expression was obtained by taking into account
relation~(\ref{eq:I(Psi)}). Substituting for \Pol{\vec{B}}{} the expression
for the dipole magnetic field at the NS surface %
$\Pol{\vec{B}}=B_0(\Uvec{r}\cos\theta+(1/2)\Uvec{\theta}\sin\theta)$ and
expressing $I$ and $\Psi$ through normalised quantities we get for the
poloidal current density in the polar cap of pulsar
\begin{equation}
  \label{eq:app-jpc-norm-1}
  \PC{j} = |\GJ{j}|\; \frac1{2}  S^\prime(\psi)
  \sqrt{1-\frac3{4}\sin^2\theta}
  \,,
\end{equation}
where $|\GJ{j}|$ is the absolute value of the Goldreich-Julian current
density in the polar cap
\begin{equation}
  \label{eq:app-jgj}
  |\GJ{j}| = \frac{B_0 \Omega}{2\pi c}\: c
  \,.
\end{equation}
For dipole magnetic field in the polar cap (see
eq.~(\ref{eq:bcond_center}))
\begin{equation}
  \label{eq:app-psi(theta)}
  \psi = \frac{\RLC}{\RNS} \sin^2\theta 
  \approx \frac{\RLC}{\RNS} \theta^2
  \,.
\end{equation}
The colatitude of the polar cap boundary is
$\PC{\theta}=1.45\times10^{-2}P^{-1/2}\sqrt{\PsiL}$, $P$ is period of
pulsar in seconds. So, $\theta^2<\PC{\theta}^2 \ll 1$ and the term
with $\sin^2\theta$ in eq.(\ref{eq:app-jpc-norm-1}) can be neglected.
From equation~(\ref{eq:app-psi(theta)}) we have relation between the
colatitude $\theta$ in the polar cap and the corresponding magnetic
flux function $\psi$
\begin{equation}
  \label{eq:app-psi(theta/thetaPC)}
  \psi = \PsiL \: \left(\frac{\theta}{\PC{\theta}}\right)^2
  \,.
\end{equation}
Substituting this relation into equation~(\ref{eq:app-jpc-norm-1}) we
get
\begin{equation}
  \label{eq:app-jpc-final}
  \PC{j} = |\GJ{j}| \; \frac1{2}  
  S^\prime 
  (
  \left( \frac{\theta}{\PC{\theta}} \right)^2 \PsiL
  )
  \,.
\end{equation}

\section[]{Energy losses}
\label{sec:App-energy-losses}

The poloidal component of Poynting flux in the magnetosphere of
aligned rotator is
\begin{equation}
  \label{eq:app-poynting}
  \Pol{\vec{P}} = \frac{c}{4\pi}\Pol{[\E \x \B]} =
  -\frac{\OmF}{c}\: I\Pol{\B}
  \,,
\end{equation}
the latter expression was obtained with help of
eqs.~(\ref{eq:B}),(\ref{eq:E}). Expressing $\Pol{\B}$ through $\Psi$,
eq.~(\ref{eq:B_components}), we have for the radial component of the
Poynting flux
\begin{equation}
  \label{eq:app-Poynting_r:general}
  P_r = - \frac{I\OmF}{c} \frac1{\varpi r}
  \left( Z\,\pd_\varpi\Psi -\varpi\,\pd_Z\Psi \right)
  \,,
\end{equation}
where $r=\sqrt{\varpi^2+Z^2}$. Energy losses trough a soling angle
$d\omega$ are $dW= - r^2 P_r\, d\omega$. The angular distribution of
energy losses are given by
\begin{equation}
  \label{eq:app-dW/dOmega:general}
  \frac{dW}{d\omega} = 
  \frac{I\OmF}{c} \frac{r}{\varpi}
  \left( Z\,\pd_\varpi\Psi -\varpi\,\pd_Z\Psi \right)
  \,.
\end{equation}
Using normalised quantities we rewrite this equation as
\begin{equation}
  \label{eq:app-dWdOmega}
  \frac{dW}{d\omega} = \frac{\Wmd}{4\pi} \; 
  S \frac{\sqrt{x^2+z^2}}{x}
  \left( z\,\pd_x\psi - x\,\pd_z\psi \right)
  \,,
\end{equation}
where $\Wmd$ is absolute value of magnetodipolar energy losses, here
defined as
\begin{equation}
  \label{eq:app-W_md}
  \Wmd \equiv \frac{\mu^2}{\RLC^4} c =
  \frac{B_0^2 \RNS^6\Omega^4}{4 c^3}
  \,. 
\end{equation}

Energy losses of aligned rotator can be obtained by integration of
equation~(\ref{eq:app-dW/dOmega:general}):
\begin{equation}
  W = \int_{4\pi} \frac{dW}{d\omega} \, d\omega =
  2 \int_0^{\Psi_\mathrm{last}} \frac{2\pi}{c}  I \OmF\: d\Psi
  \,,
\end{equation}
factor 2 appears because energy is carried away by the Poynting flux
from both hemispheres.  Using normalised quantities introduced at the
end of the section~\ref{sec:general-equation} this formula can be
rewritten as
\begin{equation}
  \label{eq:app-W}
  W = \frac{\mu^2}{\RLC^4}c \int_0^{\PsiL} S\, d\psi = 
  \Wmd \int_{0}^{\PsiL} S\, d\psi
  \,.
\end{equation}

Analytical formula for estimation of aligned rotator energy losses
could be obtained on the following way. Poloidal current $S$ for each
of obtained solutions does not deviates much from the Michel's current
function, eq.(\ref{eq:Spoloidal_Michel}). Substituting this function
into equation~(\ref{eq:app-W}) we get
\begin{equation}
  \label{eq:app-W-theory-aux}
  W \approx -\frac2{3} \PsiL^2\, \Wmd
  \,.
\end{equation}
Dependence of \PsiL{} on \XL{} we could estimate using magnetic flux
function of the dipolar field, eq.(\ref{eq:PsiDipol}). Substituting
$\psi^\mathrm{dip}(\XL) = \XL^{-1}$ into
equation~(\ref{eq:app-W-theory-aux}) we get
\begin{equation}
  \label{eq:app-W-theory}
  W \approx -\frac2{3} \XL^{-2}\,\Wmd
  \,.
\end{equation}

\section[]{The size of the region where particle inflow is possible}
\label{sec:Ux_max}

%
%
\begin{figure}
    \includegraphics[clip,width=\columnwidth]{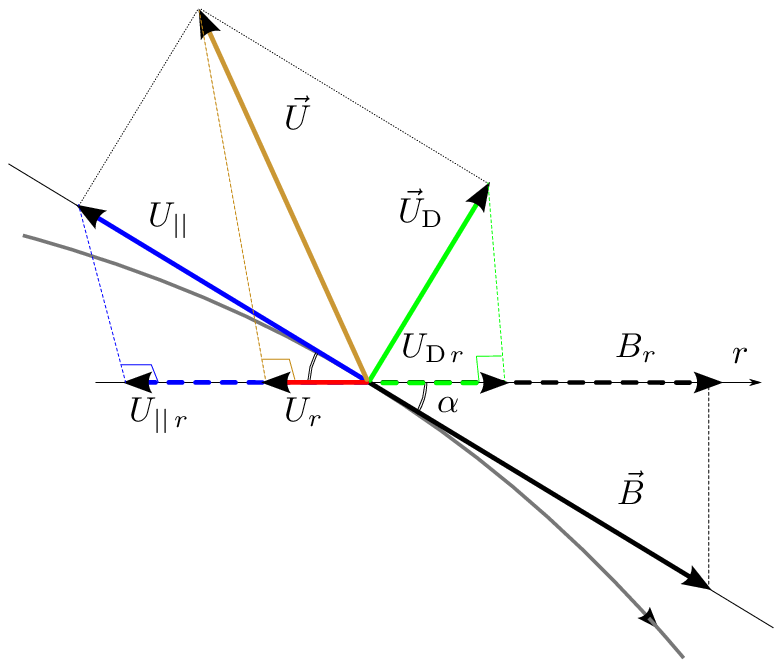}
    \caption{Particle velocity decomposition in $U_{||}$ and
      $U_\mathrm{D}$. Magnetic field line is shown by the thick curved
      line.}
  \label{fig:Ux_max}
\end{figure}

Charged particles in crossed electric and magnetic fields drift with
the velocity \Ud{} given by eq.~(\ref{eq:Udrift:Omega_B}), which can be
rewritten as
\begin{equation}
  \label{eq:app-Ud_BpBphi}
  \Ud = \frac{\OmF\varpi}{B^2} 
  (\Pol{B}^2 \Uvec{\phi} - B_\phi \vec{B}_\mathrm{pol})
  \,.
\end{equation}
In general, charged particle in the magnetosphere could have two
velocity components: one perpendicular to the magnetic field line (the
drift velocity) and an additional component along magnetic field line
$U_{||}$, see Fig.~\ref{fig:Ux_max}.  With increasing of the distance
from the NS $U_\mathrm{D}$ increases, see Fig.~\ref{fig:Main}. The
velocity of the particle $U$ can not exceed the speed of light, hence,
the parallel component of particle velocity far from NS must be also
smaller. Magnetic field lines far from NS are strongly twisted and at
some distance the particle radial velocity component will be positive,
i.e.  directed from the NS, for any direction of the parallel
component. At these distances particles can flow only \emph{from} the
NS. Let us show that the size of the domain, where particles could
flow to the NS, corresponds to the size of the light cylinder.

From Fig.~\ref{fig:Ux_max} it is evident that the radial component of the
total particle velocity is
\begin{equation}
  \label{eq:app-Ux-general}
  U_r = U_{\mathrm{D}\, r} + U_{||\, r}
  \,.
\end{equation}
The maximal value of the velocity component parallel to the magnetic
field is 
\begin{equation}
  \label{eq:app_Upar_max}
   U_{||}^\mathrm{max} \simeq \sqrt{c^2-\Ud^2}
   \,.
\end{equation}
If the radial component of the magnetic field $B_r>0$ then for
the azimuthal component of magnetic field $B_\phi=-|B_\phi|$. For
radial components of particle velocities we have
\begin{eqnarray}
  \label{eq:app_Us_varpi}
  U_{||\, r}^\mathrm{max} & = &
  - \frac{B_r}{B} U_{||}^\mathrm{max} \simeq
  - \frac{B_r}{B} c 
  \sqrt{ 1 - \left( \frac{\OmF\varpi}{c} \right)^2  \frac{\Pol{B}^2}{B^2} }\\
  U_{\mathrm{D}\, r} & = &
  \frac{\OmF\varpi}{B^2} |B_\phi| B_r
\end{eqnarray}
From this the maximal possible velocity component in radial direction
is
\begin{equation}
  \label{eq:app_Uvarpi_max}
   U_{r}^\mathrm{max} \simeq
   - \frac{B_r}{B} c 
   \left( 
     \sqrt{ 1 - \left( \frac{\OmF\varpi}{c} \right)^2 \frac{\Pol{B}^2}{B^2} }
     - \frac{\OmF\varpi}{c} \frac{|B_\phi|}{B}
   \right)
\end{equation}
Particles could flow to the NS only if $U_r^\mathrm{max}<0$,
and this is possible if
\begin{equation}
  \label{eq:app_Uvarpi_max_less_zero_cond}
  \sqrt{ 1-\left( \frac{\OmF\varpi}{c} \right) \frac{\Pol{B}^2}{B^2} } >
  \frac{\OmF\varpi}{c} \frac{|B_\phi|}{B}
\end{equation}
or
\begin{equation}
  \label{eq:app_varpy_less_RLC}
  \varpi < \frac{c}{\OmF}
  \,,
\end{equation}
i.e. only inside the light cylinder. For $B_r<0$ we get the same
result. The same restriction, namely that particles could cross the
alfvenic surface (light cylinder in our case) only in one direction,
is proved to be valid in full MHD case too \citep[see
e.g.][]{BeskinBook}.

\label{lastpage}
\end{document}